\documentclass{article}

\usepackage{titlesec}

\titleformat{\subsubsubsection}[runin]{\normalfont\normalsize\bfseries}{\thesubsubsubsection}{1em}{}
\titlespacing*{\subsubsubsection}{0pt}{3.25ex plus 1ex minus .2ex}{1.5ex plus .2ex}

\usepackage{placeins}
\usepackage{float}
\usepackage{xcolor}
\usepackage{arxiv}
\usepackage[utf8]{inputenc} 
\usepackage[T1]{fontenc}    
\usepackage{url}            
\usepackage{booktabs}       
\usepackage{amsfonts}       
\usepackage{nicefrac}       
\usepackage{microtype}      
\usepackage{lipsum}		
\usepackage{graphicx}
\usepackage{adjustbox}
\usepackage{siunitx}
\usepackage{algorithm}
\usepackage{algorithmic}
\usepackage{amsmath}
\usepackage{cite}
\usepackage[absolute,overlay]{textpos} 

\usepackage[numbers]{natbib}

\usepackage{tabularx}
\usepackage{tabularray}
\usepackage{tablefootnote}
\usepackage{booktabs}
\usepackage{longtable}
\usepackage{amsmath}

\usepackage{hyperref}
\hypersetup{
    colorlinks=true,
    linkcolor=blue,
    filecolor=magenta,      
    urlcolor=cyan,
    pdftitle={Overleaf Example},
    pdfpagemode=FullScreen,
    citecolor=blue,
    }
\begin{document}

\newpage

\title{Physics-Informed Machine Learning for Battery Degradation Diagnostics: A Comparison of State-of-the-Art Methods}

\author{Sina Navidi$^{1}$, Adam Thelen$^{2}$, Tingkai Li$^{1}$, Chao Hu$^{1,\dagger}$ \\
\\
	\normalsize $^{1}$Department of Mechanical Engineering, University of Connecticut, Storrs, CT, 06269, USA \\
	\normalsize $^{2}$Department of Mechanical Engineering, Iowa State University, Ames, IA 50011, USA \\ 
 \\
	\normalsize $^{\dagger}$Indicates corresponding author. Email: \href{mailto:chao.hu@uconn.edu}{chao.hu@uconn.edu}.
}

\date{}

\maketitle

\textbf{ACKNOWLEDGEMENTS} \\
This research was partly supported by the Iowa Economic Development Authority under the Iowa Energy Center Grant No. 20-IEC-018 and partly by the US National Science Foundation (NSF) Grant No. ECCS-2015710. Any opinions, findings, or conclusions in this paper are those of the authors and do not necessarily reflect the views of the sponsoring agencies. The authors would also like to thank Dr. Gaurav Jain and Dr. Hui Ye at Medtronic for providing the implantable-grade lithium-ion battery cells used to collect the cycling data for this study.

\vspace{35pt} 
 
\begin{abstract}

Monitoring the health of lithium-ion batteries' internal components as they age is crucial for optimizing cell design and usage control strategies. However, quantifying component-level degradation typically involves aging many cells and destructively analyzing them throughout the aging test, limiting the scope of quantifiable degradation to the test conditions and duration. Fortunately, recent advances in physics-informed machine learning (PIML) for modeling and predicting the battery state of health demonstrate the feasibility of building models to predict the long-term degradation of a lithium-ion battery cell's major components using only short-term aging test data by leveraging physics.
In this paper, we present four approaches for building physics-informed machine learning models and comprehensively compare them, considering accuracy, complexity, ease-of-implementation, and their ability to extrapolate to untested conditions. We delve into the details of each physics-informed machine learning method, providing insights specific to implementing them on small battery aging datasets. Our study utilizes long-term cycle aging data from 24 implantable-grade lithium-ion cells subjected to varying temperatures and C-rates over four years. This paper aims to facilitate the selection of an appropriate physics-informed machine learning method for predicting long-term degradation in lithium-ion batteries, using short-term aging data while also providing insights about when to choose which method for general predictive purposes.
\end{abstract}

\section{Introduction}

Monitoring the health of lithium-ion batteries over their lifetime is important for ensuring the safety and reliability of the electric vehicles and portable electronics they power. Common battery health indicators like remaining capacity and direct-current internal resistance (DCIR) can be directly measured through complete diagnostic cycles at low rates (capacity) or hybrid pulse-power characterization (HPPC) tests (resistance) \citep{barre2013review}. However, in online applications, battery capacity must be estimated because completing a full discharge cycle would significantly interrupt system operation. Battery capacity and resistance can be predicted online using reduced order algebraic models \citep{smith2021lithium, schimpe2018comprehensive, gasper2022machine}, state observer algorithms (like the extended Kalman filter and particle filters \citep{plett2004extended, hu2012multiscale, saha2009prognostics, hu2018remaining}) coupled with physics-based battery models like equivalent circuit models \citep{feng2015online, andre2011characterization, hu2012comparative} and electrochemical models \citep{moura2014adaptive, xiong2018electrochemical, bole2014adaptation}, and machine learning methods \citep{berecibar2016online, hu2015online, yang2018novel, richardson2018gaussian, shen2020deep, shen2019adeep, roman2021machine}. In cases where limited data are available, machine learning models have still been demonstrated to accurately predict battery capacity and resistance using temperature data \citep{li2022data} and partial charging data \citep{li2021feature, agudelo2023battery, deng2021data, yao2023data, li2018random}. 
Machine learning methods, such as linear regression \cite{berecibar2016online,agudelo2023battery}, support vector machines \cite{berecibar2016online}, random forest regression \cite{li2018random}, neural networks \cite{berecibar2016online, shen2019adeep, shen2020deep, roman2021machine,yao2023data},  and Gaussian process regression (GPR) or kriging \cite{yang2018novel, richardson2018gaussian, deng2021data}, have been successfully applied to estimate the capacity of batteries. These methods train a machine learning model to learn the correlation between features extracted from cell measurements (voltage, current, temperature) and the cell’s state of health (SOH).
However, none of these methods provide insight into the health of a cell's internal components, which is important for understanding the source of capacity loss and resistance increase in a cell. In real-world scenarios, battery aging occurs under wide ranges of use conditions and cases, often causing cell capacity and resistance to vary non-monotonically over a cell's lifetime \citep{gasper2022predicting}. This highlights that, even with similar capacity loss, the level of internal damage and dominant degradation mechanisms vary with operating conditions \cite{liu2020aging, thelen2022integrating}. To properly quantify cell health, more insight into the health of a cell's major internal components is required.

Three commonly reported degradation modes help to elucidate the root cause of cell capacity loss and resistance increase: they are loss of active materials on the positive and negative electrodes, abbreviated as $\mathrm{LAM_{PE}}$ and $\mathrm{LAM_{NE}}$, respectively, and loss of lithium inventory, $\mathrm{LLI}$ \citep{birkl2017degradation, thelen2022integrating, thelen2021physics, han2014comparative, dubarry2017state, tian2021electrode}. The first two modes, $\mathrm{LAM_{PE}}$ and $\mathrm{LAM_{NE}}$, describe the loss of active electrode materials, both lithiated and delithiated. Some of the causes for the loss of active materials on the positive and negative electrodes include harsh operating conditions, such as high current, which causes the electrodes to swell and crack. Other mechanisms by which the positive and negative electrodes degrade include structural disordering, gas formation, and the dissolution of metal ions into the electrolyte \cite{birkl2017degradation}. On the other hand, the $\mathrm{LLI}$ degradation mode is used to track the remaining usable lithium inventory and is closely coupled with cell capacity. The growth of a solid-electrolyte interphase (SEI) layer  is one major contributor to $\mathrm{LLI}$ \citep{attia2020revisiting}. Additionally, other lithium-consuming side reactions occur during normal cell cycling, further decreasing the lithium inventory. For example, reactions between lithium ions and impurities in the electrolyte or electrode materials can lead to the formation of solid compounds that consume lithium ions, causing a reduction in available lithium. Together, $\mathrm{LAM_{PE}}$, $\mathrm{LAM_{NE}}$, and $\mathrm{LLI}$ help better explain the component-level causes for cell-level capacity loss and resistance increase.

Generally, these three degradation modes can be experimentally verified by destructively analyzing the cell components. The loss of active electrode materials, both $\mathrm{LAM_{PE}}$ and $\mathrm{LAM_{NE}}$, can be quantified by measuring the remaining capacity of the aged electrode materials, typically achieved by cycling half-cells built using the aged electrodes. Measuring the $\mathrm{LLI}$ is more subjective since it depends on the upper voltage limit the cell is run at. To measure the $\mathrm{LLI}$ capacity, the aged full-cell can be cycled to the maximum upper voltage limit which is typically beyond the real use-case upper voltage limit. However, doing so can significantly damage the cathode material, and is therefore typically only done once the cell has been retired from operation. While these methods for quantifying the state of each degradation mode are certainly effective, they are not preferred because they require retiring the cell from the field and destroying it in the process \cite{demers2020characterization}. There is a great need to develop non-destructive degradation diagnostics methods that can be used to accurately quantify the state of the $\mathrm{LAM_{PE}}$, $\mathrm{LAM_{NE}}$, and $\mathrm{LLI}$ degradation modes online during standard cell operation.

Several non-destructive methods for estimating the state of each degradation mode have been proposed in the past. Han et al. utilized membership functions to quantify the areas under peak locations on the $dQ/dV (V)$ curve and linked them to $\mathrm{LLI}$ and $\mathrm{LAM_{NE}}$ degradation modes \cite{han2014comparative}. Birkl et al. developed a diagnostic algorithm to determine the degradation modes and verified its accuracy by reconstructing the pseudo-OCV curve of coin-cells with known levels of degradation \cite{birkl2017degradation}. Tian et al. introduced a technique for estimating electrode aging parameters, essentially quantifying the corresponding degradation modes, by selectively sampling segments of daily charging profiles and inputting them directly into a convolution neural network \cite{tian2021electrode}. Another popular approach to estimating the state of degradation modes is by simulating them using physical cell models. For example, an electrochemical model can be built to approximate the voltage response of the cell by incorporating parameters such as active material and binder mass, particle size, electrode thickness, current collector thickness, etc. \cite{lee2021robust,li2018single,reniers2019review,fan2023nondestructive,keil2020electrochemical}. Once the model is constructed, the values of the parameters can be changed to alter the model's output such that it matches the voltage data measured from an aged cell, effectively simulating the loss of active materials and lithium inventory. However, aside from being computationally intensive, a significant challenge lies in connecting these physics-based models to real-world aging phenomena, particularly in identifying the state of degradation modes based on standard aging data, such as voltage and capacity measurements obtained from reference performance tests (RPTs).

Recent advances in the field of physics-informed machine learning (PIML), particularly physics-informed neural networks (PINNs), have demonstrated significant potential in bridging the gap between physics-based models and data-driven techniques. PIML combines the strengths of physics-based models and machine learning algorithms to enhance generalization and extrapolation capabilities, even when data is limited. By integrating the principles of physics with the flexibility and scalability of machine learning, PIML approaches offer the advantage of accurately capturing complex physical phenomena with less computational complexity. In a comprehensive overview, Karniadakis et al. highlight the potential applications of PINNs across various scientific domains \cite{karniadakis2021physics}. PINNs leverage the power of neural networks to incorporate physical principles and constraints directly into the learning process, enabling more precise predictions while maintaining computational efficiency\cite{raissi2019physics}.

In the realm of battery applications, a study by Storey et al. explored different architectures for combining physics-based and machine learning models to predict battery health \cite{aykol2021perspective}. The study highlights potential architectures for combining these models while discussing the associated limitations. However, these methods have not been widely adopted by battery researchers. In a recent study, Huang et al. \cite{huang2023minn} proposed a new deep learning architecture, termed model integrated neural network, which combines a recurrent neural network and residual neural network. This model architecture was designed to incorporate physics-based equations of Newman's pseudo-two-dimensional (P2D) model into the recurrent units and compute the residuals of the partial differential equations (PDEs) in an unsupervised manner through the residual connections. They benchmarked their proposed deep learning model against a few other models consisting of a physics-based P2D model solved by an implicit differential-algebraic solver, a purely data-driven deep neural networks model, a PINN model, and a data-driven reduced-order model. In the work by Wen et al. \cite{wen2023physics}, a PINN framework was employed to estimate the rate of capacity degradation with respect to the cycle number and the current SOH of a cell being monitored. An empirical capacity degradation model, which was an improved version of Pierre Verhulst's logistic differential equation \cite{xian2013prognostics}, was utilized to capture the capacity degradation trend of a battery cell. They used a PINN architecture to fuse the prior information from the empirical model, formulated as PDEs, which took as input features with strong correlations with capacity extracted from the experimental data. Two examples of these input features are peaks of $dQ/dV (V)$ curves and charge time during the constant current charging step. Xue et al. \cite{xue2023enhanced} proposed a PINN model to enhance the computational efficiency over a single-particle model while maintaining accuracy under high C-rates. First, they employed this single-particle model to compute the distribution of lithium-ion concentration in the electrolyte given an applied current, disregarding the electrolyte dynamics. Then, the obtained results were used to train a PINN model. This PINN model incorporates a PDE solver in its loss function to solve the one-dimensional diffusion equation. The proposed model could approximate the in-electrolyte distributions of lithium-ion concentration and potential more accurately than the traditional neural network and faster than the single-particle model solved with traditional numerical methods under dynamic charging conditions. Hofmann et al. \cite{hofmann2023physics} developed what they called ``a sequential PINN'' that combined simulation data generated by a P2D model with experimental data from laboratory measurements and vehicle field data to train a neural network for SOH estimation. For the sake of consistency, we classify their PIML approach as data augmentation. When designing the input to the neural network, they utilized time-series signals such as voltage, current, temperature, SOC, and additional internal states from the simulation data (i.e., lithium-ion concentration and potential of the electrodes and electrolyte). Through feature engineering, they calculated various scalar and vectorized features for each time series signal. The final input is represented as a three-dimensional dataframe, encompassing both scalar and vectorized features. To standardize the size of input signals from diverse data sources, a binary decision feature was added, informing the network about data availability. The trained PINN model was evaluated against test datasets from simulation, laboratory, and field sources. The reported performance indicated the highest accuracy for the model when assessed on the simulation data and the lowest on the vehicle field data. In our preliminary study applying the PINN model for degradation diagnostics \cite{navidi2023physics}, we imposed half-cell model constraints on the predicted parameters of a shallow neural network. This was achieved by minimizing the difference between an experimentally measured $dQ/dV (V)$ curve and the corresponding generated curve from the half-cell model within the loss function. The proposed approach was evaluated using aging data from a long-term (3.5 years) cycling experiment on 16 implantable-grade lithium-ion cells.

The most common approach in the literature is to use physics-based models to generate training data for machine learning models. A few studies have employed equivalent circuit and electrochemical models to generate training data for deep neural networks, enabling the prediction of terminal voltage, temperature, and state of charge \cite{li2021physics,tu2023integrating, feng2020co}. However, this approach has limitations in terms of model predictive accuracy, computational efficiency, and generalizability. These limitations arise from the fact that the neural network lacks information about the discrepancies between the physical model and the actual internal state of the battery. Moreover, none of the studies have specifically addressed the problem of degradation diagnostics, except for Thelen et al., who developed a lightweight PIML model to estimate cell degradation modes without relying on late-life aging data \cite{thelen2022integrating}. By training the model with physics-based simulation data, they reduced estimation errors significantly. However, this approach requires a prior understanding of the model's inference space, which limits its ability to extrapolate to new conditions. 

In this paper, we propose two PIML approaches, namely PINN and co-kriging, specifically designed for diagnosing cell degradation in the late aging stage without relying on late-life aging data for training. To evaluate the effectiveness of these methods, we conduct a comparative study that compares the proposed approaches to two other state-of-the-art PIML techniques: data augmentation and delta learning (more specifically, delta learning with the elastic net). Notably, co-kriging also falls within the broader category of delta learning. We compare the performance of each PIML method using battery aging data obtained from a long-term ($>4$ years) cycling experiment involving 24 implantable-grade lithium-ion cells. The motivation behind this study stems from the need to comprehensively compare different PIML methods for battery degradation diagnostics. While PIML approaches have shown promise in enhancing prediction accuracy and generalizability in battery capacity estimation, a thorough investigation and comparison of their performance on battery degradation diagnostics is still lacking in the literature. By comparing PINN, co-kriging, delta learning, and data augmentation techniques, we aim to provide valuable insights into the strengths and limitations of each method. Additionally, we seek to identify the key factors that influence their performance, such as the availability of late-life aging data, the complexity of degradation processes, and the suitability of different machine learning algorithms for deployment in various situations. By addressing these objectives, our study contributes to the understanding and advancement of PIML methods for degradation diagnostics and general prediction tasks.

\section{Dataset Description}
\begin{figure}
    \centering
    \includegraphics[width=0.9\textwidth]{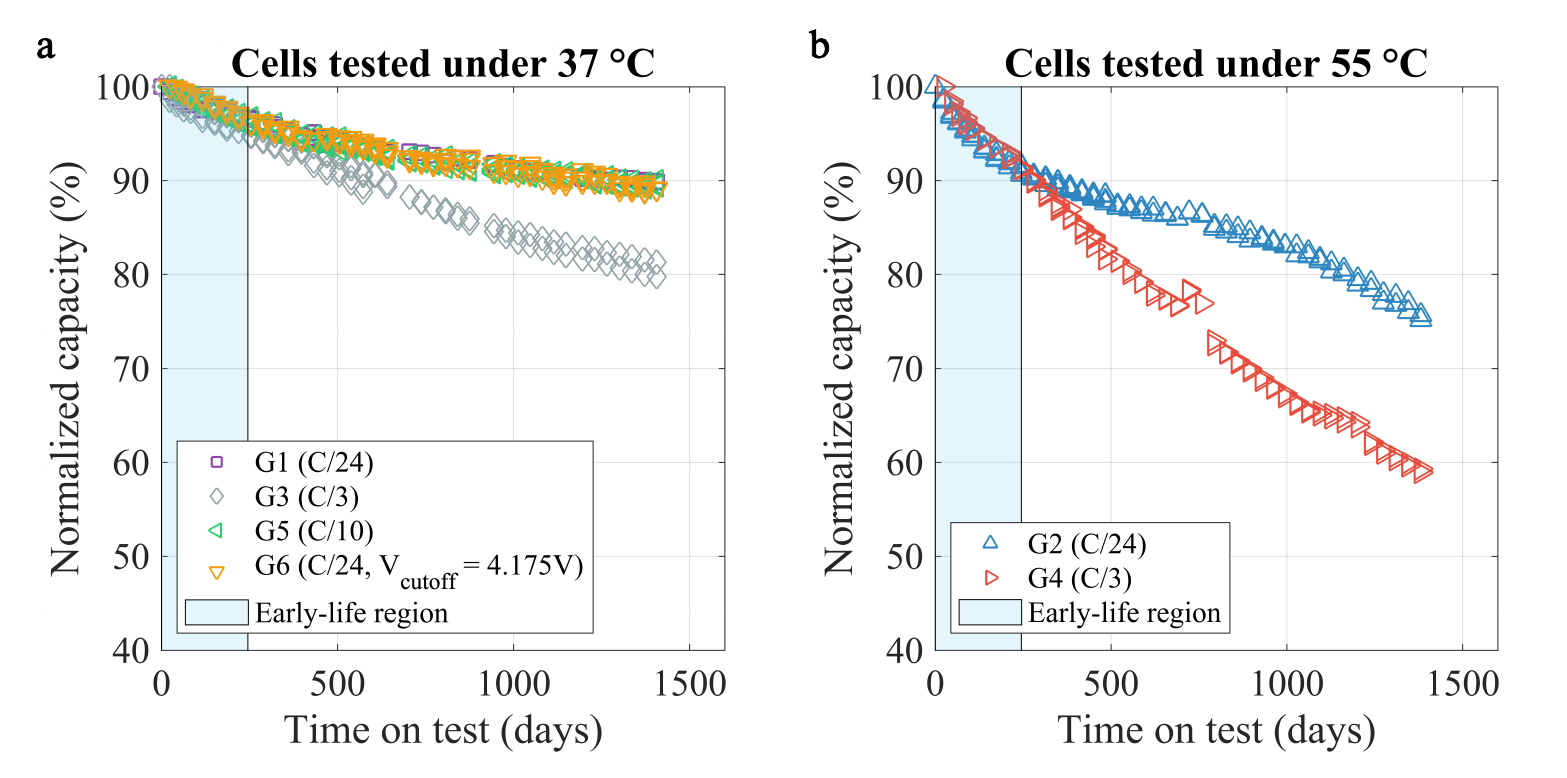}
    \caption{Overview of capacity fade curves for cells cycled under two distinct temperatures: \textbf{a}. \SI{37}{\celsius}; \textbf{b}. \SI{55}{\celsius}. The discharge rate for each group of cells appears in the legend in parentheses $(\cdot)$. The standard upper voltage cutoff ($V_{\mathrm{cutoff}}$) is set at 4.075V. The early-life region includes approximately 8 months of experimental data.}
    \label{fig:cap_fade}
\end{figure}

The battery aging dataset used in this work consists of 24 implantable-grade lithium-cobalt-oxide/graphite (LCO) cells tested in groups of four cells under six different operating conditions. The testing conditions of each group are outlined in Table \ref{tbl:conditions}. All cells are charged at C/3 to the upper voltage cutoffs, followed by constant voltage charging until the current is below C/50. Then, the cells are discharged with constant current to the lower voltage limit at 3.4 V. The aging tests were run for nearly five years in temperature-controlled thermal chambers. The capacity trajectories for this dataset can be seen in Fig. \ref{fig:cap_fade}. An RPT was performed every 2 weeks during the first 3 months and every 4 weeks thereafter. to check the cells' capacity under a common discharge rate and obtain data for half-cell model analysis and SoH estimation. During RPT, 
a common procedure has been applied across all groups. First, cells are charged to 4.075 V using a constant-current-constant-voltage protocol with a constant current at C/3 and a constant-voltage cutoff current at C/50. Second, a constant-current discharge step and a constant-current charge step, both at C/50 and followed by a half-hour rest, are applied to obtain data at a slow rate. Third, cells are discharged to 3.4 V at C/10, with a one-hour rest for every 10\% decrease in the state of charge. The temperature of thermal chambers is adjusted to \SI{40}{\celsius} before every RPT. 

\renewcommand{\arraystretch}{1.5}
\begin{table}[h!]
\caption{Summary of Aging Test Conditions}
\label{tbl:conditions}
\centering
\begin{tabular}{cccccc}
\toprule
Group & Charge rate & Discharge rate & Temperature          &Upper voltage cutoff & No. of cells     \\

\hline
G1    & C/3         & C/24           & \SI{37}{\celsius}    &4.075 V   & 4          \\
G2    & C/3         & C/24           & \SI{55}{\celsius}    &4.075 V   & 4          \\
G3    & C/3         & C/3            & \SI{37}{\celsius}    &4.075 V   & 4          \\
G4    & C/3         & C/3            & \SI{55}{\celsius}    &4.075 V   & 4         \\
G5    & C/3         & C/10            & \SI{37}{\celsius}    &4.075 V   & 4          \\
G6    & C/3         & C/24            & \SI{37}{\celsius}    &4.175 V   & 4          \\

\bottomrule
\end{tabular}
\end{table}

\renewcommand{\arraystretch}{1}

\section{Half-Cell Model}
\label{subsec:hc}

The half-cell model is a non-destructive method for quantifying three dominant degradation modes in a lithium-ion battery (i.e., $\mathrm{LAM_{PE}}$, $\mathrm{LAM_{NE}}$, and $\mathrm{LLI}$) by reconstructing a pseudo-OCV curve of the cell \citep{thelen2022integrating, birkl2017degradation}. The full-cell OCV curve is obtained by taking the potential difference between the positive and negative electrode OCV curves over the full-cell operating voltage range. At low rates ($\ll \mathrm{C}/10$), kinetic and thermal effects during charge and discharge have minimal impact on the measured OCV of the cell, making it possible to closely model full-cell OCV as the potential difference between the two individual positive and negative electrode half-cells \citep{lui2020physics, lui2021physics, birkl2017degradation, thelen2021physics}. 

\begin{figure}
    \centering
    \includegraphics[width=1\textwidth]{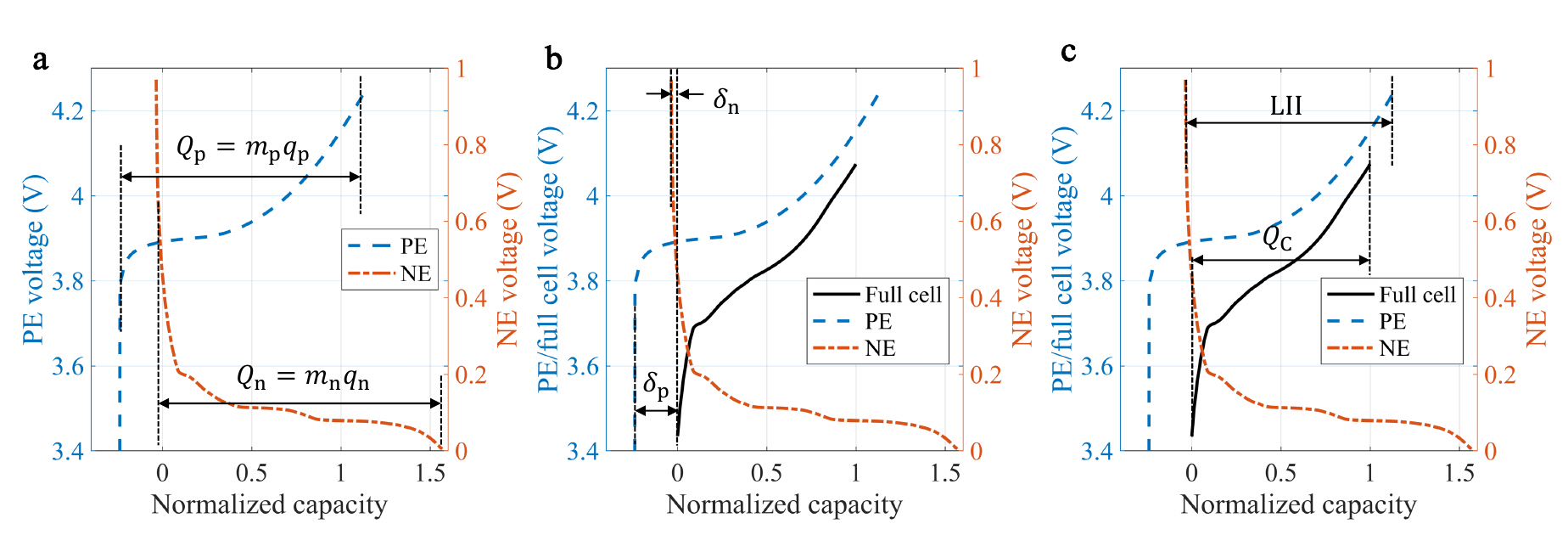}
    \caption{Overview of the half-cell model parameters. \textbf{a}. remaining active masses; \textbf{b}. slippage parameters; \textbf{c}. derived degradation parameters.}
    \label{fig:half_cell_model}
\end{figure}

The remaining active mass on the positive electrode ($m_{\mathrm{p}}$) and the negative electrode ($m_{\mathrm{n}}$) define the width of the positive and negative electrode's $QV$ curves, respectively (Fig. \ref{fig:half_cell_model}a). The remaining capacity of the electrodes is obtained by multiplying the specific capacity ($q_{\mathrm{p}}$ and $q_{\mathrm{n}}$) with the remaining active mass ($Q_\mathrm{p} = m_{\mathrm{p}}q_{\mathrm{p}}$ and $Q_\mathrm{n} = m_{\mathrm{n}}q_{\mathrm{n}}$). Then, the relative horizontal positions of these two half-cell curves are adjusted with respect to the left endpoint of the full-cell curve ($Q_{\mathrm{C}} = 0$) by two slippage parameters, $\delta_{\mathrm{p}}$ and $\delta_{\mathrm{n}}$ (Fig. \ref{fig:half_cell_model}b). The reconstructed full-cell curve is obtained by limiting the voltage range of the full-cell curve to the usable voltage window (i.e., 3.4 V to 4.075 V for RPTs). With that, we can also calculate the remaining usable capacity of the cell ($Q_{\mathrm{C}}$) and the lithium inventory indicator ($LII$),  graphically shown in Fig. \ref{fig:half_cell_model}c. $LII$ can be calculated mathematically as $LII = Q_{\mathrm{p}}-\left(\delta_{\mathrm{p}}-\delta_{\mathrm{n}}\right)$.

\subsection{Degradation Mode Quantification Using the Half-Cell Model}
\label{subsec:degradation_mode_quantification_using_the_half-cell_model}

The half-cell model is employed to quantify the true values of each degradation mode over the life of the cell. This is accomplished by manually fitting the half-cell model parameters (i.e., $m_{\mathrm{p}}$, $m_{\mathrm{n}}$, $\delta_{\mathrm{p}}$, and $\delta_{\mathrm{n}}$) to the measured C/50 voltage curves obtained from each RPT. Half-cell data from fresh electrodes was collected at C/50 to match the C-rate used in the RPT of the full-cell aging test. As the cell undergoes degradation, the widths of the positive and negative electrodes' $QV$ curves, reflecting the remaining active mass on the electrodes ($m_{\mathrm{p}}$ and $m_{\mathrm{n}}$), decrease. Simultaneously, the relative horizontal positions of the half-cell curves in relation to the left endpoint of the full-cell curve (defined by two slippage parameters, $\delta_\mathrm{p}$ and $\delta_\mathrm{n}$) also change due to the stoichiometric offset and loss of lithium inventory ($\mathrm{LLI}$). Consequently, the four parameters of the half-cell model, $m_{\mathrm{p}}$, $m_{\mathrm{n}}$, $\delta_\mathrm{p}$ and $\delta_\mathrm{n}$, influence the lengths and shapes of the two half-cell curves in specific ways, as illustrated in Fig. \ref{fig:half_cell_model} and experimentally validated in \cite{birkl2017degradation}. Thus, adjusting these parameters allows us to simulate a pseudo-OCV full-cell curve that agrees well with an experimental full-cell curve. 
To make the fitting process more accurate in reflecting the underlying cell dynamics, we also considered the fitting of width and horizontal position (corresponding to $m_{\mathrm{n}}$ and $\delta_\mathrm{n}$) and magnitude (corresponding to $m_{\mathrm{p}}$ and $\delta_\mathrm{p}$) of certain features (e.g., peaks) on the differential voltage curves ($dV/dQ (Q)$) \cite{lui2021physics,thelen2022integrating,navidi2023physics}. Manual fitting was often preferred to strike a balance between achieving a lower fitting error and leveraging the known physical meanings of certain features on the $dV/dQ (Q)$ curve, compared to automatic fitting \cite{lui2021physics}. However, this manual process is neither efficient nor practical for deployment in use cases where human intervention is not feasible (e.g., online health diagnostics inside a battery management system onboard an electric vehicle). To overcome these challenges and automate the process of degradation mode quantification, we proposed a PIML framework by leveraging both data-driven machine learning and physics-based modeling (i.e., the half-cell modeling in our earlier studies on degradation diagnostics \cite{lui2021physics,thelen2022integrating,navidi2023physics}). Furthermore, a few cells (full-cells) were removed during the aging tests to fabricate aged positive and negative half-cells whose capacity was measured through charge/discharge cycling. This effort was made to experimentally estimate the true active mass in the electrodes of each aged full-cell, serving as an experimental validation of the parameter estimates from half-cell model fitting \cite{lui2021physics}. For groups G1 and G3, cells C1 and C2 were removed for analysis at day 573, and for groups G2 and G4, cells C3 and C4 were removed at day 484. These cells were disassembled, and half-cells were fabricated from their aged electrodes. The aged half-cells were then cycled, and the measured capacities were used to confirm the loss of active materials.

\begin{figure}
    \centering
    \includegraphics[width=1\textwidth]{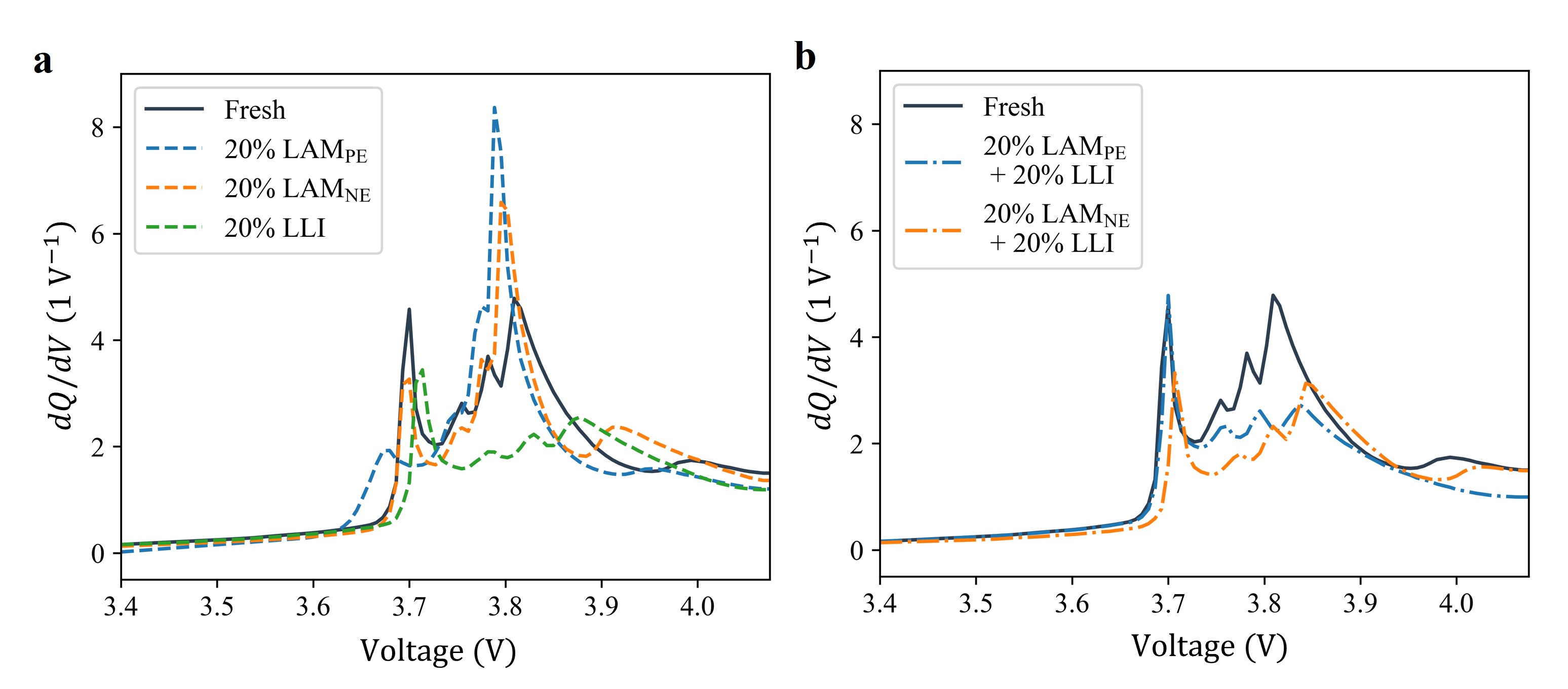}
    \caption{Simulated incremental capacity curves from the half-cell model with: \textbf{a}. one degradation mode; \textbf{b}. two degradation modes.}
    \label{fig:half_cell_sensitivity}
\end{figure}

Lithium-ion battery degradation is complex, and rarely is a single degradation mode responsible for the observed capacity fade. Researchers have long used incremental capacity analysis (ICA) and differential voltage analysis (DVA) to qualitatively assess cell degradation and identify probable degradation modes. Both ICA and DVA are useful for visualizing degradation by tracking the changes in magnitude and location of peaks and valleys that correspond to Li phase changes inside the cell. Shown in Fig. \ref{fig:half_cell_sensitivity}a, we use the half-cell model to simulate full-cell OCV and plot the incremental capacity curves ($dQ/dV (V)$) corresponding to a 20\% loss in a single degradation mode. The loss of active materials ($\mathrm{LAM_{PE}}$ and $\mathrm{LAM_{NE}}$) causes changes in the magnitude and width of the peaks in the incremental capacity curves while a loss of lithium inventory $\mathrm{LLI}$ primarily decreases the magnitude of the peaks. Additionally, another set of simulations is run to visualize the effect of two simultaneous degradation modes on the shape of the incremental capacity curve. Shown in Fig. \ref{fig:half_cell_sensitivity}b, a loss of either active material combined with a loss of lithium inventory significantly reduces the magnitude of the peaks in the incremental capacity curve, especially the larger peak around 3.8 $\si{V}$.

\subsection{Simulating Aging Data Using the Half-Cell Model}

The second way we leverage the half-cell model is for simulating cell degradation. Just as we were able to show how each degradation parameter affects a cell's incremental capacity curve in Fig. \ref{fig:half_cell_sensitivity}, we can generate many different combinations of degradation parameters and their corresponding capacity-voltage curves. Many thousands of different input-output data pairs can be simulated and used for training machine learning models. Specifically, simulation is performed by selecting a wide range of half-cell model parameters and simulating the full-cell $Q(V)$ and $dQ/dV (V)$ curves with respect to different degradation parameters from the half-cell model.

\section{Problem Definition}
Existing approaches to estimating the degradation modes present in a battery cell have a few limitations that make them impractical for deployment in the field. They generally use expensive-to-evaluate physical battery models \cite{dubarry2012synthesize, uddin2016characterising} or require collecting long-term aging test data \cite{dubarry2020big, han2014comparative, ma2018mechanism}, the latter of which makes deploying the methods extremely time and cost-intensive. 

\begin{figure} [h]
\includegraphics[width=1\textwidth]{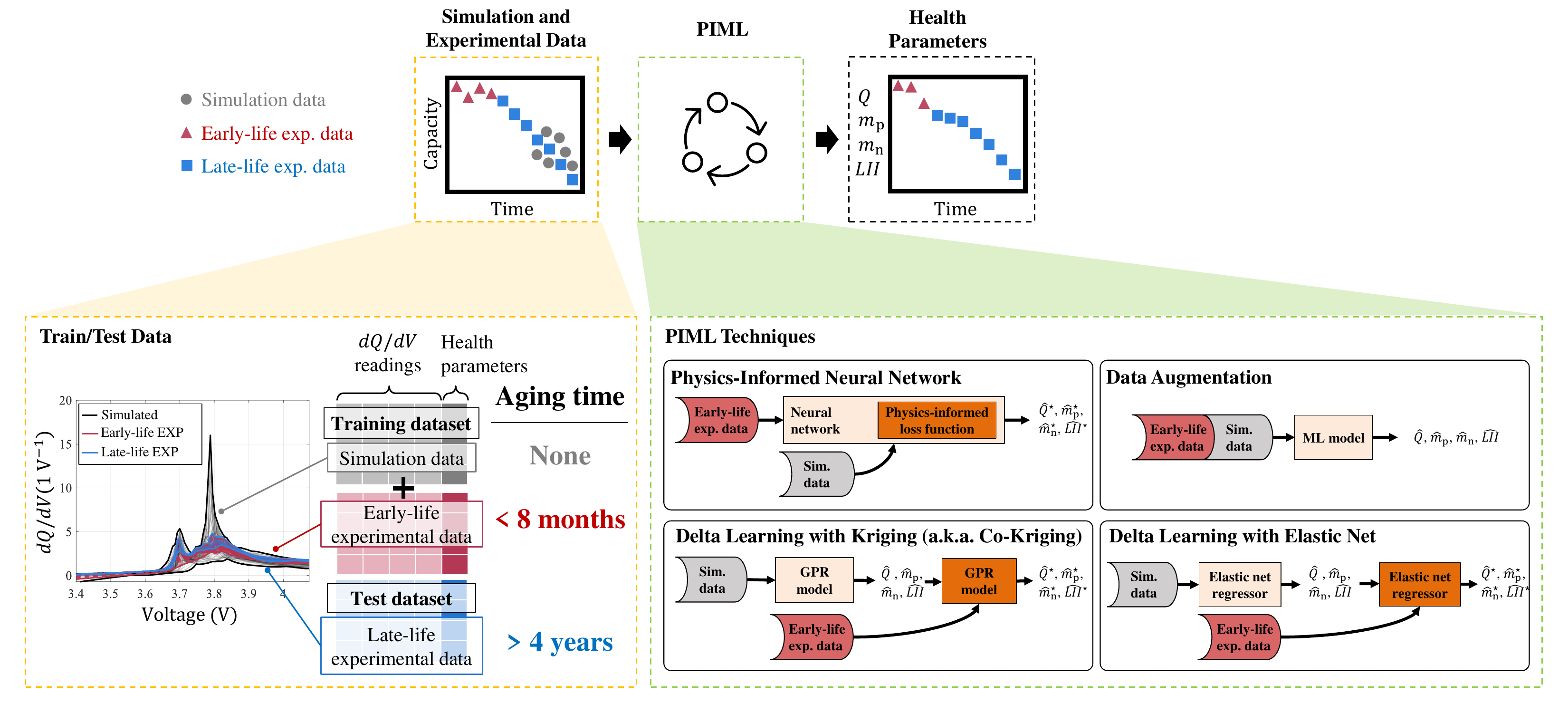}
\centering
\caption{An overview of degradation diagnostics using PIML techniques.}
\label{fig:graphical_absstract}
\end{figure}

Initial research by Thelen et al. showed that lightweight machine learning models could be trained to predict late-life degradation modes without using long-term aging data from experimental tests \citep{thelen2022integrating}. The models learned from a combination of experimental data and physics-based simulation data to estimate cell capacity and three degradation modes. We further investigate this line of research, and in this work, comprehensively study new state-of-the-art physics-informed machine learning (PIML) methods for degradation diagnostics. 

We explore the integration of physics into lightweight machine learning models through four unique approaches. Specifically, we train physics-informed machine learning models to estimate cell capacity and the state of three primary degradation modes ($\mathrm{LAM_{PE}}$, $\mathrm{LAM_{NE}}$, and $\mathrm{LLI}$) using the measured incremental capacity curve $dQ/dV (V)$ as input data. Using the differential capacity curve as input is common practice and eliminates the need for manual feature selection and enhances the applicability of the models and methods to other battery chemistries. The primary focus of this research is to conduct a comparative analysis of four distinct methods that incorporate physics-based knowledge into machine learning models. We aim to identify the most effective approach in terms of model accuracy and performance, ultimately providing insight into the optimal integration of physics concepts into machine learning frameworks for battery degradation diagnostics.

We study the degradation diagnostics problem using a framework inspired by Thelen et al. This framework leverages both early-life experimental data and simulation data from a half-cell model to construct the training dataset for our PIML methods. The process involves fine-tuning four adjustable parameters ($m_{\mathrm{p}}$, $m_{\mathrm{n}}$, $\delta_{\mathrm{p}}$, $\delta_{\mathrm{n}}$) in the half-cell model to match the reconstructed pseudo-OCV curves with those obtained from RPTs. This yields the $dQ/dV (V)$ curves, their corresponding fitted half-cell model parameters, and four health parameters (cell's capacity ($Q$) and degradation parameters ($m_{\mathrm{p}}$, $m_{\mathrm{n}}$, and $LII$)) for both the early-life and late-life stages of the aging tests. The early-life experimental data serve as the training set, while the late-life experimental data act as the testing set for our models. We employ the half-cell model to simulate data points on the  $dQ/dV (V)$ curves, along with their corresponding health parameters, for both the medium and late-life stages of each cell. These simulated data points are then incorporated into the training process alongside experimental data. The PIML methods are then trained on both the early-life experimental data and the simulated data to predict the capacity and degradation parameters for the late-life stage. Finally, we evaluate the performance of our models using late-life experimental data. Fig. \ref{fig:graphical_absstract} provides an overview of the framework.

\section{Physics-Informed Machine Learning (PIML) Techniques}

\begin{figure}
\includegraphics[width=1\textwidth]{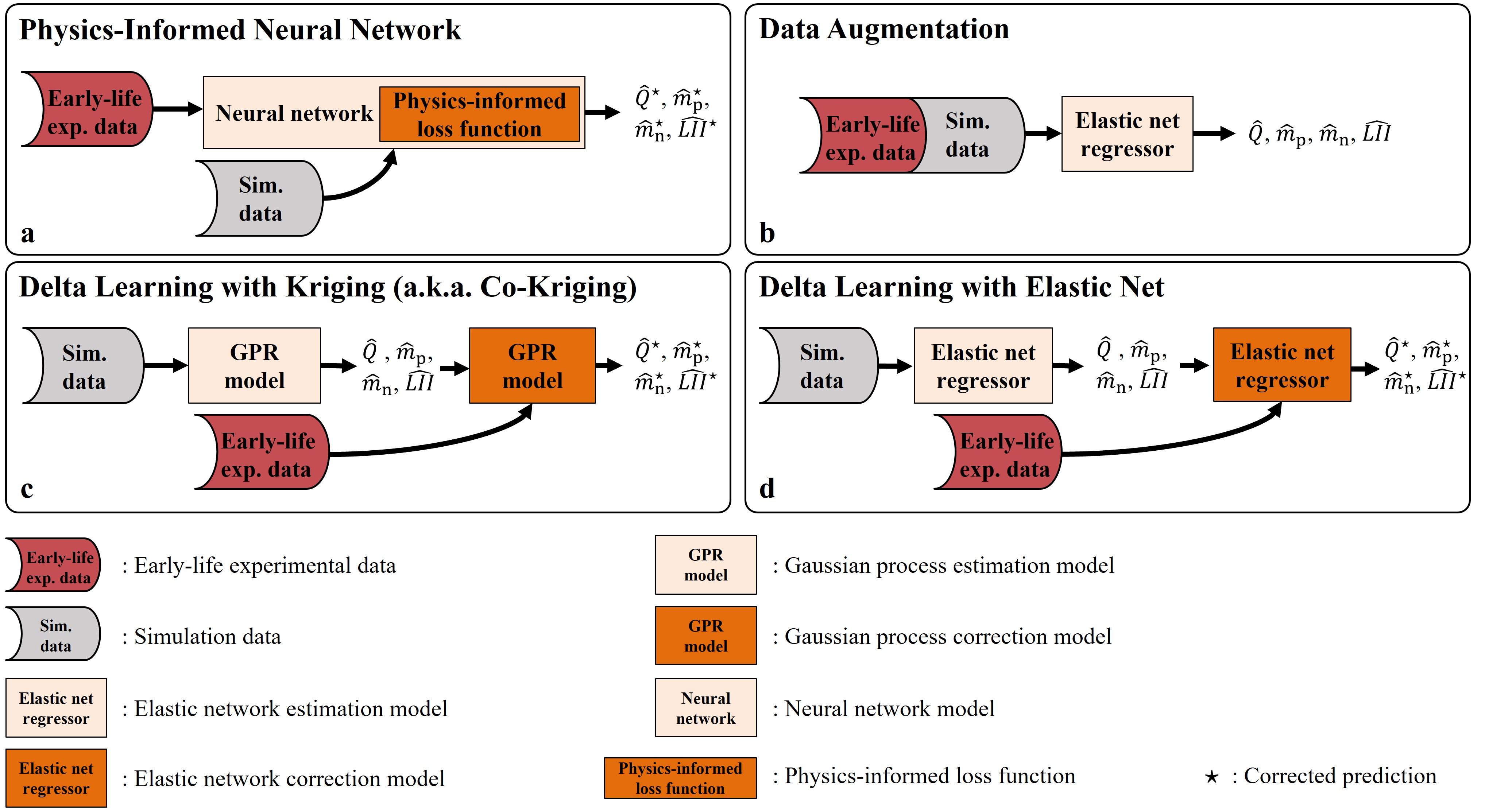}
\centering
\caption{Overview of physics-informed machine learning (PIML) techniques adopted in this comparative study.}
\label{fig:PIML methods}
\end{figure}

The following sections highlight the key characteristics of the four PIML techniques employed in our study. These PIML methods are (1) physics-informed neural networks (PINN), (2) data augmentation, (3)  delta learning with kriging (a.k.a., multi-fidelity co-kriging), and (4) delta learning with the elastic net. Fig. \ref{fig:PIML methods} provides a visual representation of these methods. A common objective among these techniques is to utilize simulation data from the late-life stage, thereby simplifying the task of extrapolating beyond the early-life experimental data toward late-life regions for the machine learning model.

\subsection{Physics-Informed Neural Network (PINN)}
\label{subsec:PINN}
\begin{figure} 

\includegraphics[width=1.0\textwidth]{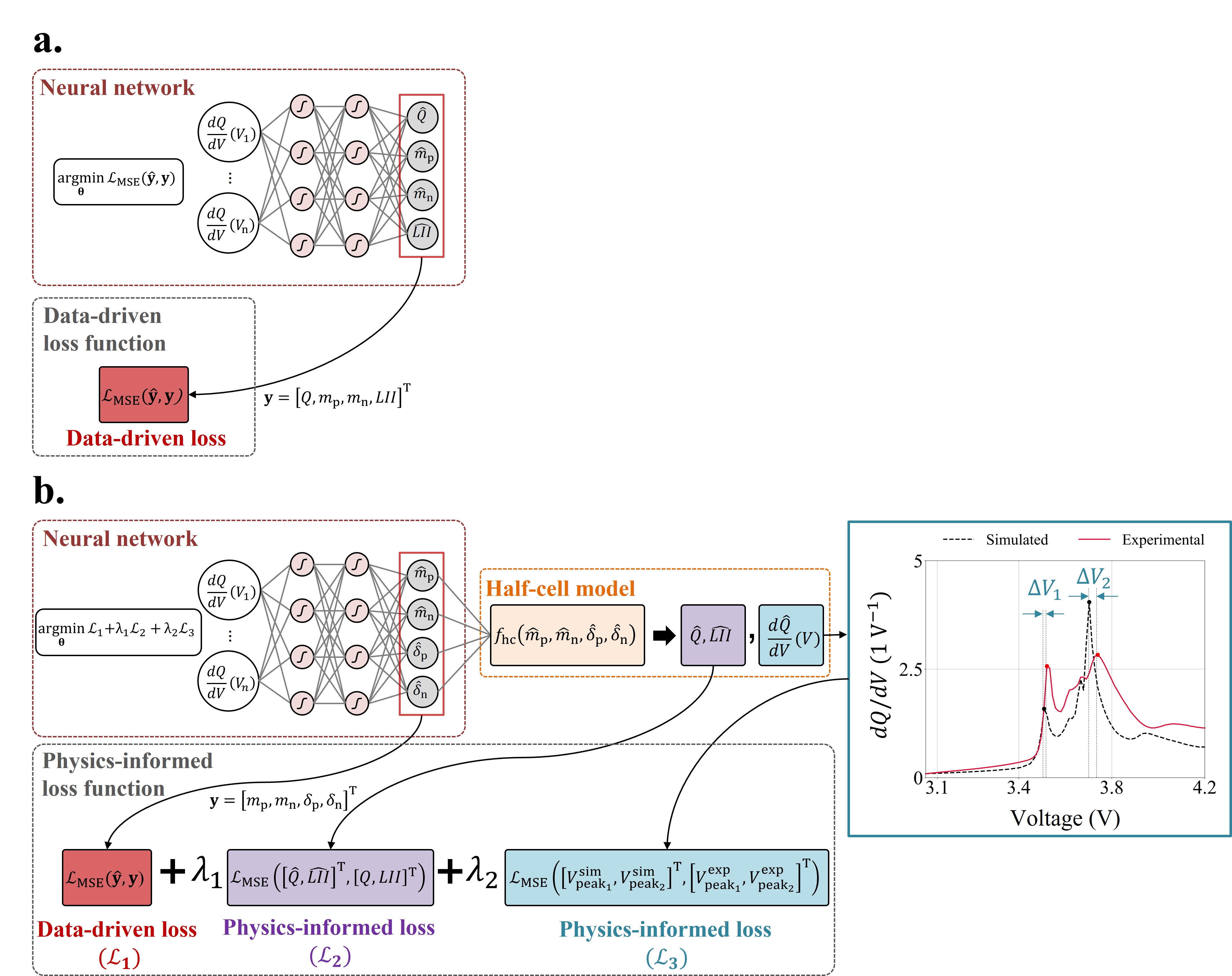}
\caption{An overview of the architectures of a baseline neural network (\textbf{a}) and PINN (\textbf{b}) for battery degradation diagnostics depicting data-driven and physics-informed loss terms.}
\label{fig:PINN}
\end{figure}

Physics-informed neural networks (PINNs) offer the advantage of fully integrating the problem's physics into their loss function, thereby enhancing generalization performance through the utilization of data and mathematical operators based on underlying physical principles \cite{karniadakis2021physics}. In our implementation, we leverage the underlying physics of the half-cell model discussed in Sec. \ref{subsec:hc} to guide the training of a neural network. This enables the network to learn the relationship between the input (a $dQ/dV (V)$ curve) and the health parameters ($Q$, $m_{\mathrm{p}}$, $m_{\mathrm{n}}$, and $LII$) throughout the battery's lifetime. As illustrated in Fig. \ref{fig:PIML methods}a, the late-life simulation data, which includes data points corresponding to the highest 20\% degradation in health parameters, is combined with the early-life experimental data to train the PINN model. This approach ensures that heavily-aged $dQ/dV (V)$ curves in the late-life region, along with their corresponding health parameters, are incorporated into the training process. To improve cell degradation estimation accuracy, particularly during the late-aging stage, we incorporate knowledge of the relationship between health parameters and $dQ/dV (V)$ into the model by introducing two physics-informed loss terms in addition to the standard data-driven loss. 
Specifically, to harness the known physics embedded in the half-cell model within the PINN model, we adopt a two-step process of first predicting the half-cell model parameters and then mapping them into the capacity and a degradation parameter, as shown in Fig. \ref{fig:PINN}b, rather than directly predicting the health parameters, as implemented in a baseline neural network (depicted in Fig. \ref{fig:PINN}a).
\begin{itemize}
  \item Initially, a shallow neural network predicts the half-cell model parameters ($m_{\mathrm{p}}$, $m_{\mathrm{n}}$, $\delta_{\mathrm{p}}$, and $\delta_{\mathrm{n}}$) for a given measurement of the $dQ/dV (V)$ curve from a fresh/aged full-cell. This neural network handles the reverse process of half-cell modeling, where the half-cell model parameters are estimated by matching the model-simulated and experimentally measured full-cell $QV$ curves. This model fitting process using the physics-based half-cell model is slow and computationally demanding due to its iterative nature, prompting us to train a neural network to perform it more efficiently. 
  \item Then, to translate these predicted half-cell model parameters into the battery capacity ($Q$) and the lithium inventory indicator ($LII$), we integrate a surrogate of the half-cell model ($f_{\mathrm{hc}}$) into the loss function. In defining the physics-informed loss terms, we leverage the forward mapping of the physics-based half-cell model, which proves to be straightforward and fast.
\end{itemize}

As depicted in Fig. \ref{fig:PINN}, a shallow network is utilized to establish a mapping of the 100 points on the early-life experimental and late-life simulation curves of $dQ/dV (V)$ to the four half-cell model parameters ($\delta_{\mathrm{p}}, \delta_{\mathrm{n}}, m_{\mathrm{p}}, m_{\mathrm{n}}$). The standard data-driven loss is calculated using the predicted half-cell model parameters and their corresponding true values obtained from early-life experimental and late-life simulated data. Let $\mathbf{y} = [m_{\mathrm{p}}, m_{\mathrm{n}}, \delta_{\mathrm{p}}, \delta_{\mathrm{n}}]^\mathrm{T}$ be the vector of true values and $\mathbf{\hat{y}} = [\hat{m_{\mathrm{p}}}, \hat{m_{\mathrm{n}}}, \hat{\delta_{\mathrm{p}}}, \hat{\delta_{\mathrm{n}}}]^\mathrm{T}$ be the vector of predicted parameters. The MSE over both the early-life experimental points and late-life simulated points can be calculated as:

\begin{equation}
\mathcal{L}_1 = \mathcal{L}_{\mathrm{MSE}}(\mathbf{\hat{y}}, \mathbf{y}) = \frac{1}{{N_{\mathrm{exp}} + {N_{\mathrm{sim}}}}} \left( \sum_{i=1}^{N_{\mathrm{exp}}} \| \mathbf{y}_i - \mathbf{\hat{y}}_i \|^2 + \sum_{j=1}^{N_{\mathrm{sim}}} \| \mathbf{y}_j - \mathbf{\hat{y}}_j \|^2 \right)
\end{equation}

where $N_{\mathrm{exp}}$ and $N_{\mathrm{sim}}$ denote the numbers of data points for the early-life experimental points and late-life simulated points, respectively; $\mathbf{y}_i$ and $\mathbf{\hat{y}}_i$  represent the true and predicted parameter vectors for the $i$-th early-life point; and $\mathbf{y}_j$ and $\mathbf{\hat{y}}_j$ represent the $j$-th element of the true health parameters vector and predicted vector for the late-life points. 

Within the loss function, a half-cell surrogate model, denoted as $\mathrm{f}_{hc}$, is integrated to establish a mapping between the predicted half-cell model parameters and the battery capacity ($Q$) and the lithium inventory indicator ($LII$). This integration ensures that the values of battery capacity and lithium inventory remain consistent with the physics-based half-cell model by verifying that the network-predicted half-cell model parameters lead to simulated curves with endpoints that closely match those of the true fitted curves. Consequently, a second loss term is generated to measure the difference between the capacity($\hat{Q}$) and $\hat{LII}$ degradation parameter obtained by passing the predicted half-cell model parameters from the network into the half-cell surrogate model and the true ones. Let $\mathbf{U} = [Q, LII]^\mathrm{T}$ be the vector of true values and $\mathbf{\hat{U}} = [\hat{Q}, \hat{LII}]^\mathrm{T}$ be the vector of outputs from the half-cell surrogate. The $\mathcal{L}_2$ loss can be calculated as:

\begin{equation}
\mathcal{L}_2 =\mathcal{L}_{\mathrm{MSE}}(\mathbf{\hat{U}}, \mathbf{U}) = \frac{1}{N} \sum_{k=1}^N \|\mathbf{\hat{U}}_k - \mathbf{U}_k\|^2
\end{equation}

where $N = N_{\mathrm{exp}} + N_{\mathrm{sim}}$ represents the total number of data points corresponding to the early-life experimental and late-life simulated measurements. Here, $\mathbf{\hat{U}}_k$ and $\mathbf{U}_k$ represent the $k$-th element of the predicted and true parameters vector, respectively. The use of a half-cell surrogate model prevents any disruption in the gradient computation process that would occur if the half-cell model were directly used within the loss function. This ensures that the predicted parameters remain connected to the computational graph in each epoch, enabling the computation of the gradient of the network output with respect to the inputs. 

Another valuable piece of information that can be integrated into the physics-based loss function is the difference between the peak positions (voltages) observed in the simulated and experimental $dQ/dV (V)$ curves. Typically, two main peaks can be observed on the $dQ/dV (V)$ curve. The first peak at the lower voltage corresponds to the intercalation of lithium ions into the electrode material during discharge. As lithium ions are released from the anode during discharge, they intercalate into the cathode's active material, leading to an increase in capacity. The second peak corresponds to the electrochemical reactions occurring at the cathode, such as the deintercalation of lithium ions from the cathode material. The shape, intensity, and position of these peaks provide valuable insights into the electrochemical behavior and performance of the battery \cite{weng2013board}. As we have constrained two of the parameters in the loss function ($Q$ and $LII$), we also aim to constrain the other two predicted parameters, which are the positive and negative active masses ($m_{\mathrm{p}}$ and $m_{\mathrm{n}}$). Since the phase transition peaks in the $dQ/dV (V)$ curve are sensitive to the values of $m_{\mathrm{p}}$ and $m_{\mathrm{n}}$ \cite{lui2021physics}, by constraining the position of these peaks in the loss function, we can ensure that the resulting active mass parameters are consistent with the half-cell model constraints. To incorporate this information, we introduce an additional loss term that calculates the difference between the positions of the two major peaks in the experimental $dQ/dV (V)$ curves and their corresponding simulated counterparts. To achieve this, we extend the half-cell surrogate model to map the predicted half-cell model parameters ($\delta_{\mathrm{p}}, \delta_{\mathrm{n}}, m_{\mathrm{p}}, m_{\mathrm{n}}$) by the network to the positions of the two major peaks in the $dQ/dV (V)$ curves. A peak detection algorithm is employed, utilizing criteria such as prominence, distance, height, and width, to identify the major peaks on the curves. The identified peaks are then sorted in descending order based on their corresponding y-values, and the first two peaks are selected for comparison. Let $\mathbf{\hat{V}}_{\mathrm{peak}}^{\mathrm{sim}} = [V_{\mathrm{peak}_1}^{\mathrm{sim}}, V_{\mathrm{peak}_2}^{\mathrm{sim}}]^\mathrm{T}$ represent the vector of peak positions on the simulated $dQ/dV (V)$ curve, and $\mathbf{V}_{\mathrm{peak}}^{\mathrm{exp}} = [V_{\mathrm{peak}_1}^{\mathrm{exp}}, V_{\mathrm{peak}_2}^{\mathrm{exp}}]^\mathrm{T}$ represent the vector of actual peak positions corresponding to the training set. The loss function $\mathcal{L}_{\mathrm{MSE}}$ is utilized to measure the mean squared error between these vectors, yielding the loss term $\mathcal{L}_3$:

\begin{equation}
\mathcal{L}_3 = \mathcal{L}_{\mathrm{MSE}}(\mathbf{\hat{V}}_{\mathrm{peak}}^{\mathrm{sim}}, \mathbf{V}_{\mathrm{peak}}^{\mathrm{exp}}) = \frac{1}{2} \left( \|\mathrm{\hat{V}}_{\mathrm{peak}_1}^{\mathrm{sim}} - \mathrm{V}_{\mathrm{peak}_1}^{\mathrm{exp}}\|^2 + \|\mathrm{\hat{V}}_{\mathrm{peak}_2}^{\mathrm{sim}} - \mathrm{V}_{\mathrm{peak}_2}^{\mathrm{exp}}\|^2 \right)
\end{equation}

Finally, to ensure that the resulting loss terms are on the same scale, we compute a weighted summation of these terms, as depicted in Fig. \ref{fig:PINN}. 
\begin{equation}
\mathcal{L}_{\mathrm{total}} = \mathcal{L}_1 + \lambda_1  \mathcal{L}_2 + \lambda_2  \mathcal{L}_3
\end{equation}

where $\lambda_1$, and $\lambda_2$ are positive scalar values that control the relative weight of each physics-informed loss term, and they are empirically determined. The weighted summation ensures that the different loss terms contribute proportionally to the overall loss function, allowing for a balanced optimization process. The custom total loss term directly reflects the higher-level objective we want the model to minimize, rather than relying solely on the loss between the true and predicted values of the network. Essentially, the training process involves updating the network weights in a manner that minimizes the customized total loss term value computed in each iteration.

\subsection{Data Augmentation}
Data augmentation is a technique used to enhance the accuracy of machine learning models by expanding the training data. One approach involves leveraging a simple and efficient physics-based simulation model to generate additional input/output samples that closely resemble the original dataset. Augmenting samples into the training data increases the dataset's size and diversity, which enables the model to learn from a wider range of examples \cite{van2001art}.

In this study, we enhance the training dataset by incorporating simulation data from the half-cell model, initially comprising only early-life battery degradation data from the cycling experiment. The simulation data is sampled from the entire design space of health parameters, including data from the late-life stage. This augmented dataset provides more information about the future degradation path of the experimental cells than what can be learned solely from the early-life experimental data. Consequently, it facilitates a more accurate estimation of late-life health parameters. For a visual representation of this approach, refer to Fig. \ref{fig:PIML methods} c. It is worth noting that the augmentation process can be implemented in different ways, depending on the specific selection of simulated degradation data to be combined with the early-life experimental data. One approach is to add simulation data points that are filtered to include only the highest degradation values for each of the health parameters: $Q$, $m_{\mathrm{p}}$, $m_{\mathrm{n}}$, and $LII$. By focusing on these higher degradation levels, the augmented dataset may offer additional insights into the extreme trends and behavior exhibited by the battery cells.

\subsection{Delta Learning}
Delta learning is a technique where a secondary model is trained to correct the errors or biases of a primary model. The secondary model, known as the delta model or correction model, is specifically designed to account for discrepancies between the predictions of the primary model and the observed experimental data. By incorporating the learned corrections from the delta model, the accuracy and performance of the primary model can be improved \cite{attia2020revisiting}.
In this study, as depicted in Fig. \ref{fig:PIML methods} b and Fig. \ref{fig:PIML methods}c, we implement delta learning using two different machine learning methods: GPR, also known as kriging, which leads to the formulation of multi-fidelity co-kriging, and the elastic net, which corresponds to a conventional form of delta learning. The key difference between these two delta learning approaches lies in the unique probabilistic formulation of co-kriging for computing the posterior distribution or final prediction, in contrast to the deterministic formulation of the elastic net.

\subsubsection{Delta Learning with Kriging (a.k.a Multi-Fidelity Co-Kriging)}
\label{subsec:Co-kriging}
GPR, also known as kriging, is a probabilistic machine learning method used to estimate an unknown mapping function $f(\cdot)$. It samples a random realization of this function from a Gaussian process, representing a distribution over a collection of functions. A Gaussian process is defined by its mean function, $\mu(\cdot)$, and covariance function (kernel), $k(\cdot, \cdot)$. The mean function captures the expected value or general trend of the data being modeled, while the kernel describes the covariance between the function outputs at two different inputs. In GPR, a Gaussian process prior is assumed on the underlying mapping function, $f(\cdot)$, using a prior covariance function. This covariance function incorporates hyperparameters and a mean function, which can be chosen based on existing knowledge or assumptions. Various types of kernels, such as linear, quadratic, or general basis functions, can be considered, depending on the prior understanding of the problem. An example of a common kernel function is provided in Eq. (\ref{Matern}) which is used in this study as well. It incorporates both distance-based and smoothness-based components to capture the underlying patterns in the data. 
\begin{equation}
\label{Matern}
k_{\mathrm{Matern}}(\mathbf{x}, \mathbf{x'}) = \sigma^2 \left(1 + \frac{\sqrt{2\nu}}{l}\|\mathbf{x} - \mathbf{x'}\|\right) \exp\left(-\frac{\sqrt{2\nu}}{l}\|\mathbf{x} - \mathbf{x'}\|\right)
\end{equation}
where $\sigma^2$ represents the variance, controlling the overall variability of the Gaussian process. The parameter $\nu$ determines the smoothness of the function, with half-integer values (e.g., 1/2, 3/2, 5/2) providing a flexible and smooth representation of the data. The length scale parameter $l$ determines the range over which the correlation between data points decreases. Smaller values of $l$ result in a more localized correlation, while larger values yield a smoother and more global correlation. The term $\|\mathbf{x} - \mathbf{x'}\|$ denotes the Euclidean distance between data points $x$ and $x'$. The exponential function $\mathrm{exp}(\cdot)$ attenuates the correlation as the distance between data points increases.

Several methods based on Gaussian processes have been developed to combine information from models with different levels of fidelity, depending on the complexity of the relationships between these fidelities. In the field of engineering design, a commonly used approach is co-kriging, which combines multiple models providing different levels of accuracy or fidelity. It assumes a linear relationship between these models, allowing for the integration of their predictions.
Building upon the two-level multi-fidelity modeling proposed in \cite{kennedy2000predicting}, we consider high-fidelity data (i.e., early-life experimental data) denoted as $\mathbf{Y}_{\mathrm{H}} = (\mathbf{y}_{\mathrm{H}}^{(1)},...,\mathbf{y}_{\mathrm{H}}^{(N_\mathrm{H})})^\mathrm{T}$ at locations $\mathbf{X}_{\mathrm{H}} = (\mathbf{x}_{\mathrm{H}}^{(1)},...,\mathbf{x}_{\mathrm{H}}^{(N_\mathrm{H})})^\mathrm{T}$, and low-fidelity data (i.e., simulation data spanning the entire lifetime of the battery cells) denoted as $\mathbf{Y}_{\mathrm{L}} = (\mathbf{y}_{\mathrm{L}}^{(1)},...,\mathbf{y}_{\mathrm{L}}^{(N_{\mathrm{L}})})^\mathrm{T}$ at locations $\mathbf{X}_{\mathrm{L}} = (\mathbf{x}_{\mathrm{L}}^{(1)},...,\mathbf{x}_{\mathrm{L}}^{(N_\mathrm{L})})^\mathrm{T}$. Here, $\mathbf{y}_{\mathrm{H}}^{(i)}, \mathbf{y}_{\mathrm{L}}^{(i)} \in \mathbb{R}^4$ (since we are predicting cell's capacity and three degradation parameters), and $\mathbf{x}_{\mathrm{H}}^{(i)}, \mathbf{x}_{\mathrm{L}}^{(i)} \in \mathbb{R}^{d}$, where $d$ represents the number of points on the input $dQ/dV (V)$ curve. By concatenating the data from both fidelities, i.e., $\mathbf{X} = \{{\mathbf{X}_{\mathrm{H}},\mathbf{X}_\mathrm{L}\}}$ and $\mathbf{Y} = {\{\mathbf{Y}_{\mathrm{H}},\mathbf{Y}_{\mathrm{L}}\}}$, we can construct a multivariate Gaussian process using co-kriging based on the auto-regressive model \cite{barajas2018multivariate}.
\begin{equation}
\label{CK}
f_{\mathrm{H}}(\mathbf{x}) = \rho f_{\mathrm{L}}(\mathbf{x}) + f_{\Delta}(\mathbf{x}),
\end{equation}
where $\mathbf{x}$ represents the input vector at which the prediction is made, $f_{\mathrm{L}}(\mathbf{x})$ and $f_{\mathrm{H}}(\mathbf{x})$ are the low-fidelity and high-fidelity Gaussian processes that model the simulation and experimental data, respectively. $f_{\Delta}(\mathbf{x})$ is a Gaussian process that captures the difference between $f_{\mathrm{H}}(\mathbf{x})$ and $\rho f_{\mathrm{L}}(\mathbf{x})$, and $\rho$ is a {regression parameter}. In this approach, it is assumed that:
\begin{equation}
\label{CK assumption}
Cov \{f_{\mathrm{H}}(\mathbf{x}), f_{\mathrm{L}}(\mathbf{x'}) | f_{\mathrm{L}}(\mathbf{x})\} = 0, \quad \forall \mathbf{x} \neq \mathbf{x'}, \mathbf{x}, \mathbf{x'} \in \mathbb{R}^{d},
\end{equation}

To implement this formulation, as illustrated in Fig. \ref{fig:PIML methods}b, we first train a GPR model to construct the $f_L(\cdot)$ using the dataset $\{{\mathbf{X}_{\mathrm{L}}, \mathbf{Y}_{\mathrm{L}}}\}$. Next, we calculate the discrepancy matrix $\mathbf{Y}_{\mathrm{\Delta}} = \mathbf{Y}_{\mathrm{H}} - \rho f_{\mathrm{L}}(\mathbf{X_{\mathrm{H}}})$, where $f_{\mathrm{L}}(\mathbf{X_{\mathrm{H}}})$ represents the mean values of $f_{\mathrm{L}}(\cdot)$ at the corresponding locations common to those of $\mathbf{X_{\mathrm{H}}}$. Finally, we train another GPR model to obtain $f_{\Delta}(\cdot)$ by training it on the dataset $\{{\mathbf{X}_{\mathrm{H}}, \mathbf{Y}_{\mathrm{\Delta}}}\}$. The final high-fidelity prediction at a new location $\mathbf{x}_{*}$, i.e., a vector of input points on a $dQ/dV (V)$ curve, consists of a posterior distribution $\hat{\mathbf{y}}(\mathbf{x}_{*}) \sim \mathcal{N}(\hat{\mathbf{\mu}}(\mathbf{x}_{*}), \hat{\mathbf{s}}^2(\mathbf{x}_{*}))$, with posterior mean and variance given by the following equations \cite{yang2019physics}.


\begin{equation}
\label{posterior mean}
\hat{\mathbf{\mu}}(\mathbf{x}_{*}) = \rho \hat{\mu}_\mathrm{L}(\mathbf{x}_{*}) + \hat{\mu}_{\Delta}(\mathbf{x}_{*}) + \mathbf{c}_{\Delta}(\mathbf{x}_{*}) \mathbf{C}_{\Delta}(\mathbf{X}_{\mathrm{H}},\mathbf{X}_{\mathrm{H}})^{-1} (\mathbf{Y}_{\Delta}- \hat{\mu}_{\Delta}(\mathbf{X}_{\mathrm{H}})),
\end{equation}

\begin{equation}
\label{posterior variance}
\hat{\mathbf{s}}^2(\mathbf{x}_{*}) = \rho^2 {{\sigma}_{\mathrm{L}}^2}(\mathbf{x}_{*}) + {{\sigma}_{\Delta}^2}(\mathbf{x}_{*}) - \mathbf{c}(\mathbf{x}_{*})^\mathrm{T} \mathbf{C}^{-1} \mathbf{c}(\mathbf{x}_{*}),
\end{equation}

where $\hat{\mu}_{\mathrm{L}}{(\mathbf{x}_{*})}$ is obtained based on $f_{\mathrm{L}}(\cdot)$, and $\hat{\mu}_{\mathrm{\Delta}}{(\mathbf{X}_{\mathrm{H}})}=(\hat{\mu}_{\Delta}^{(1)},...,\hat{\mu}_{\Delta}^{(N_\mathrm{H})})^\mathrm{T}$ and $\mathbf{C}_{\mathrm{\Delta}}(\mathbf{X}_{\mathrm{H}}, \mathbf{X}_{\mathrm{H}})$ is obtained based on constructed $f_{\Delta}(\cdot)$. Also, 
${{\sigma}_{\mathrm{L}}^2}(\mathbf{x}_{*})= k_{\mathrm{L}}(\mathbf{x}_{*}, \mathbf{x}_{*})$, and ${{\sigma}_{\mathrm{\Delta}}^2}(\mathbf{x}_{*})= k_{\Delta}(\mathbf{x}_{*}, \mathbf{x}_{*})$, and the vector of covariances ($\mathbf{c}$) for a given input $\mathbf{x}_{*}$, is expressed by Eq. (\ref{cov vector}).

\begin{equation}
\label{cov vector}
\mathbf{c}(\mathbf{x}_{*}) = \begin{bmatrix} \rho\mathbf{c}_{\mathrm{L}}(\mathbf{x}_{*}) \\ \mathbf{c}_{\mathrm{H}}(\mathbf{x}_{*}) \end{bmatrix} = \begin{bmatrix} \begin{bmatrix} \rho k_{\mathrm{L}}(\mathbf{x}_{*}, \mathbf{x}_{\mathrm{L}}^{(1)}), \ldots, \rho k_{\mathrm{L}}(\mathbf{x}_{*}, \mathbf{x}_{\mathrm{L}}^{(N_\mathrm{L})}) \end{bmatrix}^\mathrm{T} \\ \begin{bmatrix} k_{\mathrm{H}}(\mathbf{x}_{*}, \mathbf{x}_{\mathrm{H}}^{(1)}), \ldots, k_{\mathrm{H}}(\mathbf{x}_{*}, \mathbf{x}_{\mathrm{L}}^{(N_\mathrm{H})}) \end{bmatrix}^{\mathrm{T}} \end{bmatrix},
\end{equation}

where $k_{\mathrm{H}}(\mathbf{x}, \mathbf{x}^{'})= \rho^2 k_{\mathrm{L}}(\mathbf{x}, \mathbf{x}^{'}) + k_{\Delta}(\mathbf{x}, \mathbf{x}^{'})$ \cite{forrester2008engineering} and $\mathbf{c}_{\Delta}$ is obtained as $\mathbf{c}_{\Delta}(\mathbf{x}_{*}) = \mathbf{c}_{\mathrm{H}}(\mathbf{x}_{*}) - \rho ^{2} \mathbf{c}_{\mathrm{L}}(\mathbf{x}_{*})$. Additionally, the covariance matrix representing the covariances across all observations is denoted as $\mathbf{C}$ and is defined by Eq. (\ref{cov matrix}).

\begin{equation}
\label{cov matrix}
{\mathbf{C}} = \begin{bmatrix}
    \mathbf{C}_{\mathrm{L}}(\mathbf{X}_{\mathrm{L}}, \mathbf{X}_{\mathrm{L}})  &  \rho \mathbf{C}_{\mathrm{L}}(\mathbf{X}_{\mathrm{L}}, \mathbf{X}_{\mathrm{H}}) 
    \\
    \rho \mathbf{C}_{\mathrm{L}}(\mathbf{X}_{\mathrm{H}}, \mathbf{X}_{\mathrm{L}})  & \rho^2 \mathbf{C}_{\mathrm{L}}(\mathbf{X}_{\mathrm{H}}, \mathbf{X}_{\mathrm{H}}) + \mathbf{C_{\mathrm{\Delta}}}(\mathbf{X}_{\mathrm{H}}, \mathbf{X}_{\mathrm{H}})
\end{bmatrix},
\end{equation}

where $\mathbf{C_{\mathrm{L}}}$ represents the covariance matrix based on the kernel of $f_{\mathrm{L}}(\cdot)$, and $\mathbf{C_{\mathrm{\Delta}}}$ represents the covariance matrix based on the kernel of $f_{\mathrm{\Delta}}(\cdot)$, we can identify their hyperparameters, along with $\rho$, by assuming parameterized forms for these kernels. This identification process is achieved through the maximization of the log-marginal likelihood term shown in Eq. (\ref{log-likelihood term}). 

\begin{equation}
\label{log-likelihood term}
\ln L ~ = -\frac{1}{2}({\mathbf{Y}} - {\mathbf{\mu}})^\mathrm{T} {\mathbf{C}}^{-1} ({\mathbf{Y}} - {\mathbf{\mu}}) - \frac{1}{2} \ln |{\mathbf{C}}| - \frac{N_\mathrm{H} + N_\mathrm{L}}{2} \ln (2\pi).
\end{equation}

The steps of the implemented algorithm are summarized in Algorithm \ref{algorithm_1}.

\begin{algorithm}[H]
\caption{Two-fidelity co-kriging algorithm}
\label{algorithm_1}
\begin{algorithmic}
    \STATE \textbf{Step 1:} Use a GPR model trained on simulation data generated from the half-cell model across the entire design space of health parameters to construct ${f}_{\mathrm{L}}(\cdot)$ and compute $\hat{\mu}_{\mathrm{L}}{(\mathbf{X}_{\mathrm{L}})}=(\hat{\mu}_{\mathrm{L}}^{(1)},...,\hat{\mu}_{\mathrm{L}}^{(N_\mathrm{L})})^\mathrm{T}$ and $\mathbf{C}_{\mathrm{L}}(\mathbf{X}_{\mathrm{L}}, \mathbf{X}_{\mathrm{L}})$.

    \STATE \textbf{Step 2:} Use $f_{\mathrm{L}}(\cdot)$ to estimate $\hat{\mathbf{y}}_{\mathrm{H}}$ at locations corresponding to $\mathbf{X}_{\mathrm{H}}$: $f_{\mathrm{L}}(\mathbf{X}_{\mathrm{H}})$.
    
    \STATE \textbf{Step 3:} Calculate the discrepancy matrix as follows: $\mathbf{Y}_{\mathrm{\Delta}} = \mathbf{Y}_{\mathrm{H}} - \hat{\mu}_{\mathrm{L}}(\mathbf{X}_{\mathrm{H}})$. This is based on the reasoning that $\hat{\mu}_{\mathrm{L}}(\mathbf{X}_{\mathrm{H}})$ is the most probable observation of the Gaussian process $f_{\mathrm{L}(\cdot)}$.
    
    \STATE \textbf{Step 4:} Choose a specific kernel function $k_{\Delta}(\cdot,\cdot)$ (Matern kernel in this work), and train a GPR model using the dataset $\{\mathbf{X}_{\mathrm{H}}, \mathbf{Y}_{\mathrm{\Delta}}\}$ to construct $f_{\mathrm{\Delta}}(\cdot)$. Then compute $\hat{\mu}_{\mathrm{\Delta}}{(\mathbf{X}_{\mathrm{H}})}=(\hat{\mu}_{\Delta}^{(1)},...,\hat{\mu}_{\Delta}^{(N_\mathrm{H})})^\mathrm{T}$ and $\mathbf{C}_{\mathrm{\Delta}}(\mathbf{X}_{\mathrm{H}}, \mathbf{X}_{\mathrm{H}})$.
    
    \STATE \textbf{Step 5:} The posterior mean and variance for any given input, $\mathbf{x}_{*}$, are computed by Eq. (\ref{posterior mean}) and Eq. (\ref{posterior variance}), respectively. 
\end{algorithmic}
\end{algorithm}

\subsubsection{Delta Learning with Elastic Net}
In another implementation of delta learning (see Fig. \ref{fig:PIML methods}d), we employ a simple elastic net model trained on simulation data to estimate the health parameters of cells in the late-life stage. However, since the simulation data does not precisely reflect the heavily aged $dQ/dV (V)$ curves observed in the experimental tests, the estimation model alone may produce inaccurate predictions. To overcome this limitation, we introduce a corrector elastic net model which is trained using early-life degradation data, and its purpose is to capture the "delta" or prediction bias of the first model for the light-degradation region. This learned bias can be extended to effectively correct the outputs in the late-life stage. During the training phase, the corrector model leverages the available early-life experimental data to learn the prediction bias of the estimation model. In the testing phase, both the estimation model and the corrector model work in conjunction. The final prediction is generated by combining the output of the estimation model with the correction provided by the corrector model. 
It is worth noting that the implementation structures of delta learning and co-kriging are quite similar in this study. The main difference lies in the choice of estimation and corrector models. In co-kriging, we utilized two GPR models, whereas in delta learning, we employed two elastic net models. Broadly speaking, we can consider the co-kriging implementation in this study as a special case of delta learning.

\section{Results}

In this section, we present a detailed comparative analysis of four PIML methods for battery cell degradation diagnostics. Our goal is to assess each model's performance in comparison to its data-driven baseline and explore specific analyses for the novel approaches, PINN and co-kriging. We start by evaluating the overall performance of each PIML model relative to its data-driven counterpart, providing insights into their predictive accuracy. We also investigate the impact of varying training sample sizes on performance and assess the extrapolation capabilities of PINN and co-kriging. Finally, we conduct sensitivity analysis by investigating the effects of changing model parameters and structure. Through this analysis, we offer a comprehensive view of the strengths and limitations of each PIML method in battery cell degradation diagnostics.

\subsection{Training and Test Setup}
\begin{figure} 
\includegraphics[width=1.0\textwidth]{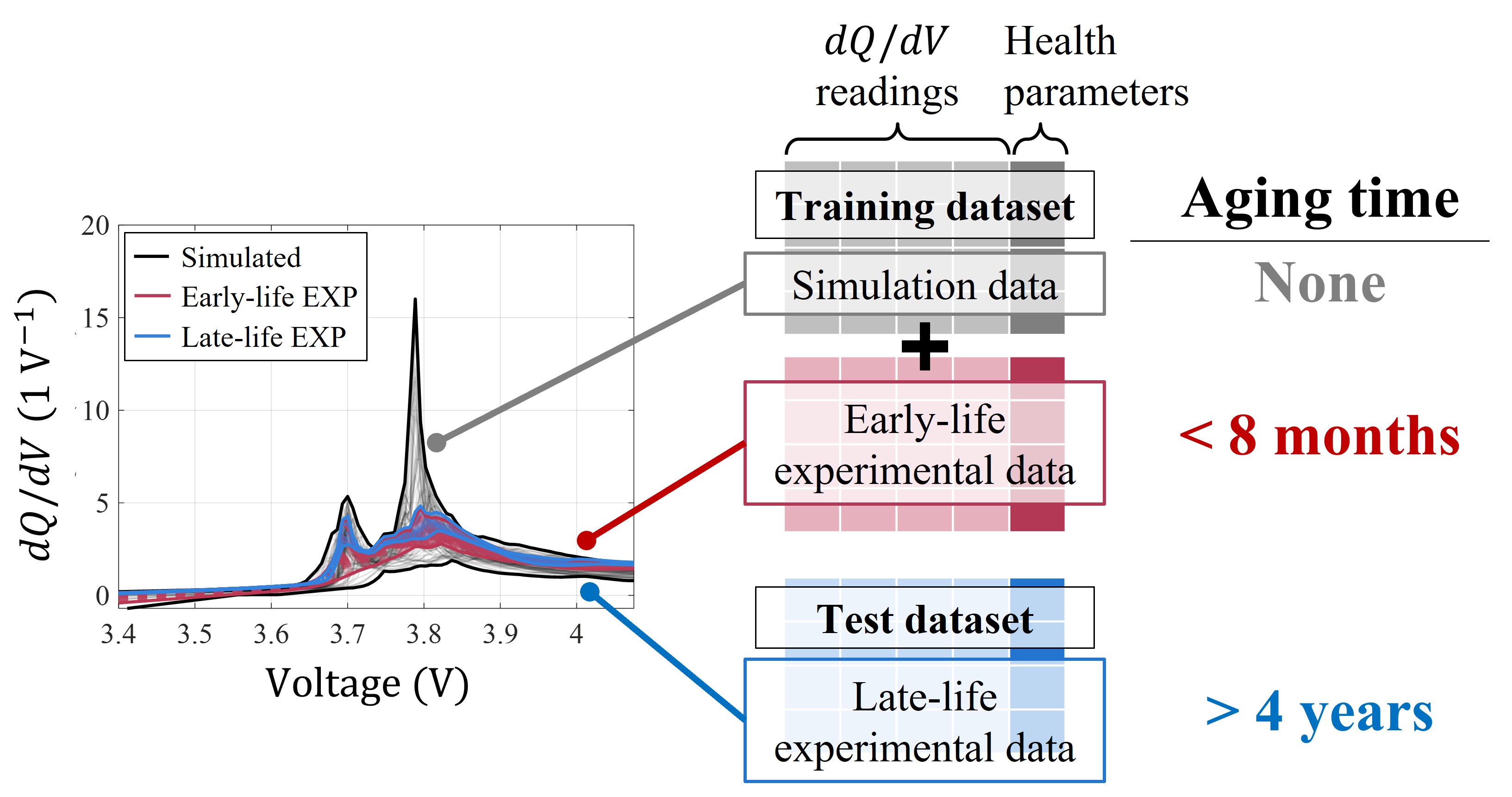}
\centering
\caption{Training and test setup for PIML approaches.}
\label{fig:setup}
\end{figure}

To assess the effectiveness of the PIML techniques, we trained each model using the initial 10 characterization points, representing approximately seven months of experimental aging data. Subsequently, we compared the performance of each PIML model with its respective data-driven baseline counterpart, which was solely trained on early-life experimental data points. The key distinction between the PIML methods and their baseline counterparts lies in how they integrate the late-life simulation data into the training process. These methods use different approaches to incorporate this data, aiming to achieve improved extrapolation performance in the late-life stage. The general setup for train/test split is shown in the Fig. \ref{fig:setup}. 

To comprehensively evaluate the performance of the PIML models, we conducted a four-fold cross-validation study. This approach ensures that the entire experimental dataset, comprising 24 cells, is thoroughly assessed while considering potential variations across different subsets of the data. In our study, we divided the dataset into four mutually exclusive folds, with each fold including one battery cell from each group in the test set. Table \ref{table:points} summarizes the numbers of training and test points for each fold.



\begin{table}[h]
    \centering
    \caption{Numbers of training points in 4-fold cross-validation}
    \label{table:points}
    \begin{tabular}{lcccc}
        \toprule
        Method & Fold & Training Points & Testing Points \\
        \midrule
        PINN & 1,2 & 212 & 200 \\
        & 3,4 & 212 & 192 \\
        \midrule
        Data Augmentation & 1,2 & 212 & 200 \\
        & 3,4 & 212 & 192 \\
        \midrule
        Delta Learning & 1,2 & 964 & 200 \\
        & 3,4 & 964 & 192 \\
        \midrule
        Baseline Methods & 1,2 & 180 & 200 \\
        & 3,4 & 180 & 192 \\
        \bottomrule
    \end{tabular}
\end{table}

In the following subsections, we provide details about the parameter settings implemented in the training process for PINN and co-kriging, two newly proposed approaches.

\subsubsection{PINN}
We opted for a simplified neural network architecture in our PINN approach, comprising two hidden layers. This choice reflects our emphasis on simplicity and considers the size limitation of the available data, which makes a complex neural network susceptible to overfitting. By employing a simpler network architecture with fewer parameters, we intentionally limit the model's ability to memorize noise present in the data, thereby promoting improved generalization. To further mitigate the risk of overfitting, we implemented an early stopping mechanism. This mechanism involved monitoring the validation loss during training and terminating the training process when the validation loss started to increase. Additionally, we incorporated cross-validation techniques and integrated simulated data from the late-life stage into the training process. These strategies collectively contribute to enhancing the generalization capability of the PINN model and serve as effective safeguards against overfitting, differentiating it from a conventional neural network model. During model training, we utilized the Adam optimization algorithm to iteratively update network weights. The rectified linear unit (ReLU) function was used as the activation function throughout the training process, defined as:
\begin{equation}
\label{ReLU}
ReLU(x) = \mathrm{max}(0, x)
\end{equation}
The hyperparameters of the network are summarized in Table \ref{table:hyper}.

\renewcommand{\arraystretch}{1.5}
\begin{table}[h!]
\caption{Neural network hyperparameters}
\label{table:hyper}
\centering
\begin{tabular}{cc}
\toprule
Hyperparameter                & Value  \\
\hline
Number hidden of layers               & 2        \\
Max epoch                             & 1000-2000       \\
Number of hidden neurons              & 30         \\
Learning rate                         & 0.005-0.01       \\
Batch size                            & 200            \\
Trainable parameters                  & 2274            \\
\bottomrule
\end{tabular}
\end{table}
\renewcommand{\arraystretch}{1}
In the testing phase, the trained network operates independently, removing the need for the physics-informed loss function. When combined with the half-cell model, the trained network assesses degradation levels at various stages of cell life. Essentially, the trained network takes over the manual half-cell fitting process by predicting the half-cell model parameters.

\subsubsection{Co-Kriging}
Training a kriging (GPR) model with a specific kernel involves determining several parameters to effectively capture the underlying patterns in the data. The Matern kernel, which is used in this study, is a versatile kernel function that can be adjusted with a few hyperparameters to fit the characteristics of the data. The key parameters that should be determined when training a GPR model with a Matern kernel are summarized in the Table \ref{table:CK_par}.


\begin{table}[h]
    \centering
    \caption{Key parameters for training GPR models with Matern kernel for each fold}
    \label{table:CK_par}
    \begin{tabular}{lccc}
        \toprule
        Parameter & Estimation Model & Correction Model \\
        \midrule
        Length scale ($l$) & 5258.770 & 1.000 \\
        Signal variance ($\sigma^2$) & 0.015 & 0.006-0.008 \\
        Smoothness parameter ($\nu$) & 3/2 & 3/2 \\
        \bottomrule
    \end{tabular}
\end{table}

\subsection{Error Metrics}
For each fold, we calculated an average test error of each individual PIML model for each of the four health parameters ($Q$, $m_{\mathrm{p}}$, $m_{\mathrm{n}}$, and $LII$). These errors measured the average differences between the predicted parameter values and the corresponding ground truth values, which were the degradation parameters obtained through fitting the half-cell model to experimental full-cell curves (see Sec. \ref{subsec:degradation_mode_quantification_using_the_half-cell_model}). We computed a normalized average of the individual health parameter test errors across the four folds using the following root mean squared percentage error (RMSPE) formula:
\begin{equation}
\label{RMSPE}
RMSPE_t = \sqrt{\frac{1}{\sum_{k=1}^{4} N_k} \sum_{k=1}^{4} \sum_{i=1}^{N_k} \left(\frac{{\hat{y}_{t,i} - y_{t,i}}}{{y_{t,i}}}\right)^2} \times 100 \%
\end{equation}
Here, $N_k$ represents the number of test samples, and the subscript $t$ denotes the $t$-th degradation parameter. $\hat y_{ti}$ and $y_{ti}$ correspond to the predicted value and true value, respectively, for the $t$-th health parameter at the $i$-th test point. Note that our earlier study on degradation diagnostics relevant to this work used ``RMSE\%" in place of ``RMSPE" \citep{thelen2022integrating}.

\subsection{Overall Performance Comparison}

\begin{figure}
\includegraphics[width=1\textwidth]{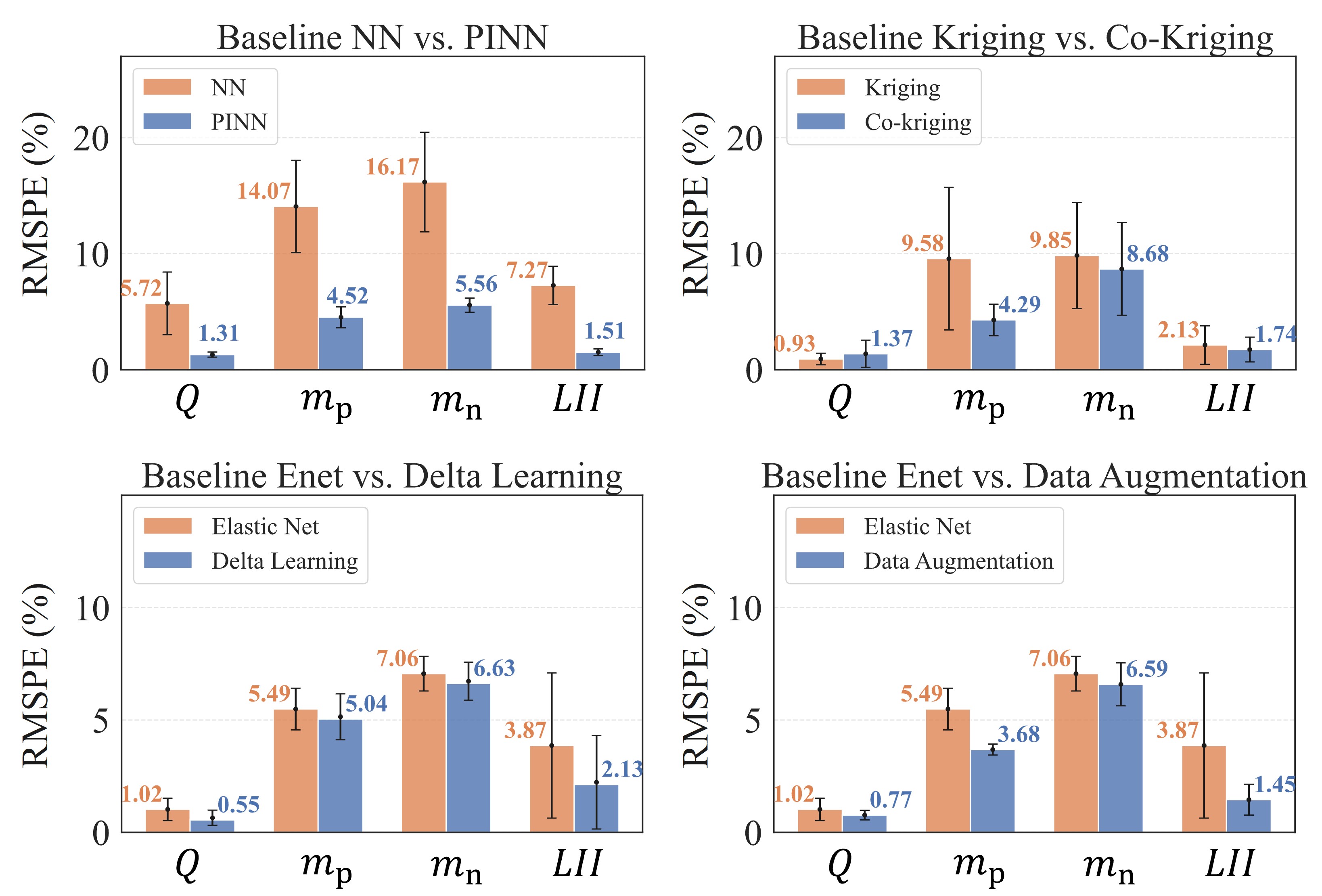}
\centering
\caption{Overall performance comparisons between physics-informed ML models and purely data-driven baselines.}
\label{fig:Overall_Performance}
\end{figure}

The results, illustrating the average test errors for individual health parameters, can be found in Fig. \ref{fig:Overall_Performance}. 
As depicted in Fig. \ref{fig:Overall_Performance}, all the PIML methods exhibited improved error rates compared to their data-driven counterparts. It is noteworthy that the degradation trends of the negative and positive active mass parameters ($m_{\mathrm{p}}$ and $m_{\mathrm{n}}$) were generally more complex and challenging to predict compared to the capacity and lithium inventory. The degradation of electrode active materials involves intricate physical and chemical processes, such as solid-state diffusion, phase transformations, and mechanical stresses. These processes interact in a highly nonlinear and complex manner, making it challenging to model and predict their combined effects accurately. Despite these inherent difficulties, the PIML methods demonstrated a high level of accuracy in predicting these parameters. This highlights the effectiveness of the PIML approaches in capturing the underlying degradation patterns and enhancing the prediction capabilities for the more intricate parameters.

Specifically, the average test error of the PINN demonstrated a remarkable improvement compared to its baseline neural network model with the same network structure. The baseline neural network displayed a broader range of errors, implying a higher degree of variability in its predictions' behavior across different runs. It's important to note that the increased width of the error bars reflects greater variability in performance across different training runs, rather than solely indicating elevated predictive uncertainty within the trained models. In contrast, the PINN consistently delivered improved performance in each fold with a smaller range of errors across multiple runs. The superior performance of the PINN compared to the baseline neural network can be attributed to the PINN's enhanced extrapolation capability, as showcased in Fig. \ref{fig:extrapolation_PINN}, stemming from two key factors. First, the PINN leveraged a concept known as observational bias, as discussed in the research by Pateras et al. \cite{pateras2023taxonomic}. This means that during the learning (model training) process, a deliberate physical bias was introduced. This bias can be implemented through the use of physics-informed data structures, in the context of either simulation or observation. This was achieved by incorporating simulated late-life stage degradation data alongside early-life experimental data during training. This approach minimized a sampling error, providing a more comprehensive representation of the underlying data distribution. In contrast, the baseline neural network relied solely on early-life experimental data. Second, the PINN adopted an inductive bias \cite{pateras2023taxonomic} approach by integrating the half-cell model within its loss function (Fig. \ref{fig:PINN}b). This integration allowed the PINN to predict the half-cell model parameters rather than directly predicting the health parameters. This intermediate prediction step ensured the alignment of the final predicted health parameters with the constraints imposed by the half-cell model. These constraints were incorporated by passing the network-predicted parameters to the half-cell model and introducing two physics-informed loss terms to the loss function. These terms were minimized through the training process. In contrast, the baseline neural network predicted health parameters directly from early-life experimental data without imposing the constraints informed by the physics-based half-cell model, as shown in Fig. \ref{fig:PINN}a.
To ensure a fair comparison, we performed ten cross-validation runs for each model and computed the mean and standard deviation of the RMSPE across these runs. Additionally, while training these models, we regularly assessed their performance using a validation dataset, which is a distinct subset of the training dataset, including data that the model hasn't encountered during its training phase. An early stopping mechanism was also integrated into the training process. Particularly, if the validation loss fails to exhibit improvement over a specified number of epochs (50 in this case), we halt the training prematurely. This strategy aims to prevent the model from memorizing noise inherent in the training data. By doing so, we ensure that the model version used for evaluation possesses stronger generalization capabilities.

Regarding co-kriging, we observed significant improvements in the error rates for the positive active mass ($m_{\mathrm{p}}$) and capacity ($Q$) parameters. Additionally, the error rates for the negative active mass and lithium inventory parameters were also lower compared to the purely data-driven GPR (kriging) baseline model. It is important to note that the GPR baseline model was trained exclusively on the early-life experimental data and did not incorporate the simulation data from the half-cell model.

We benchmarked data augmentation and delta learning techniques against a simple elastic net model (baseline), trained using only the early-life experimental data. Notably, data augmentation exhibited superior performance to delta learning when trained on data encompassing the entire simulated parameter range, which includes both medium and high-degradation regions. Specifically, the data augmentation technique showed substantial improvement, particularly for the positive active mass parameter ($m_{\mathrm{p}}$), compared to the baseline elastic net model. On the other hand, the error rates for the other three parameters were relatively close for both delta learning and data augmentation. 

In the evaluation of baseline models, the elastic net outperformed both the regular neural network and GPR (kriging) model in terms of overall accuracy. The baseline neural network exhibited the lowest overall accuracy, which was somewhat expected given the limited amount of training data available for network training. Notably, the baseline kriging model demonstrated excellent predictive capabilities, particularly in estimating capacity ($Q$) and lithium inventory ($LII$). However, when it came to predicting the active mass parameters ($m_{\mathrm{p}}$ and $m_{\mathrm{n}}$), the elastic net emerged as the most accurate baseline model.

\subsection{Performance Under Different Training Sample Sizes}
\begin{figure} 
\includegraphics[width=0.9\textwidth]{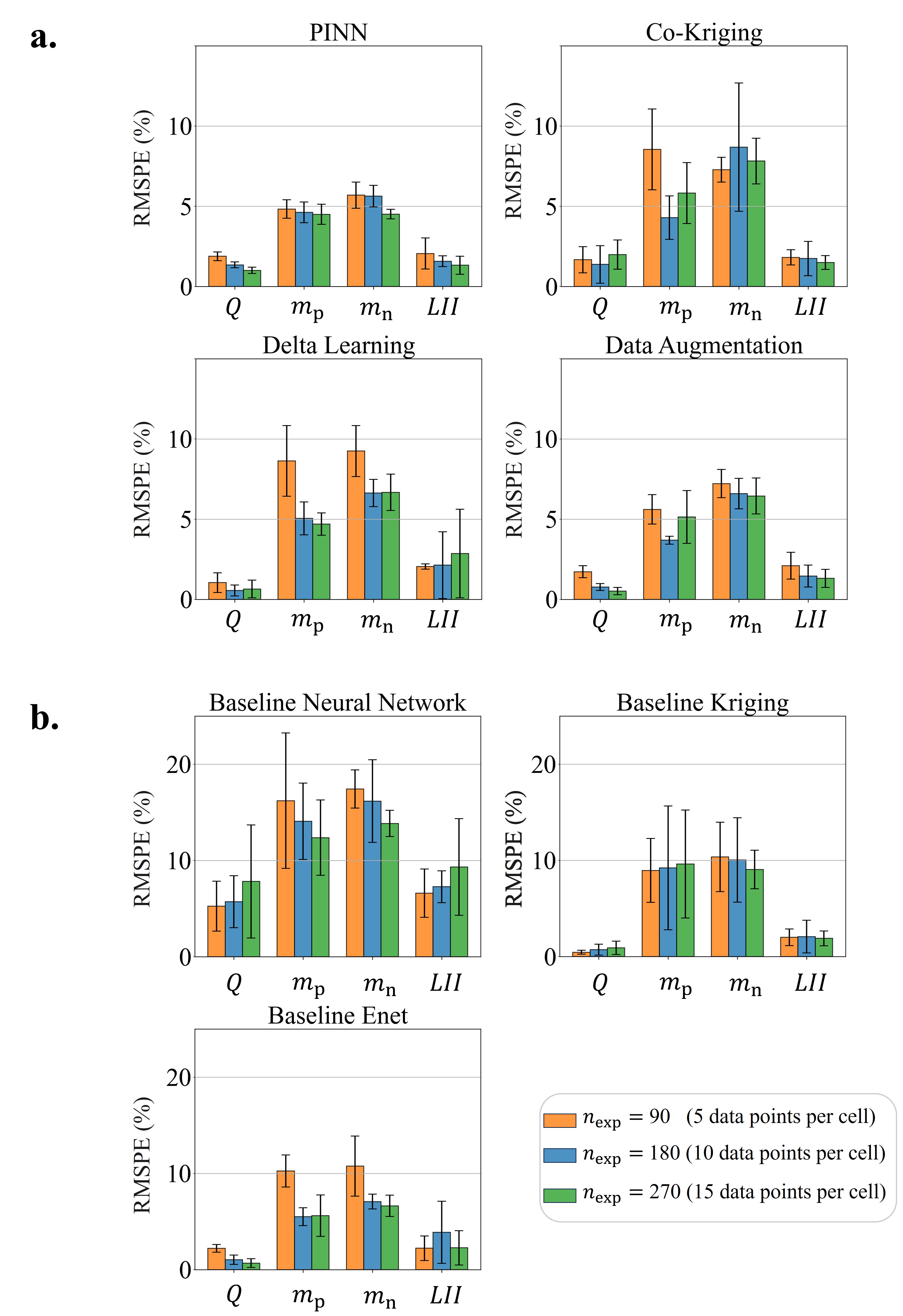}
\centering
\caption{Impact of training sample size on the performance of \textbf{a}. PIML models \textbf{b}. baseline models ($n_{\mathrm{exp}}$ represents the total number of experimental data points).}
\label{fig:Training_size}
\end{figure}

Here, we examined the effect of including a different number of early-life experimental data points ($n_{\mathrm{exp}}$) in the training process on the performance and error rates of PIML models. To carry out this analysis, we kept the test set fixed, following the same methodology as before. In each fold, the test set consisted of the cell from each group corresponding to that fold. However, we varied the number of experimental characterization points in the training set; specifically, we explored training scenarios involving 5, 10, and 15 data points per training cell. This range of scenarios resulted in a total training point count varying from 90 to 270 across different iterations. To provide context, the number of data points per cell corresponds to different durations of cycle aging data. With 5 data points per training cell, we consider 3 months of cycle aging data. With 10 data points per training cell, we extend the duration to 8 months, and with 15 data points per training cell, we utilize 12 months of cycling data. These variations allow us to investigate the impact of including different amounts of early-life data on the model's performance. The comparative results of this analysis are summarized in Fig. \ref{fig:Training_size}a for PIML methods considered. By examining the relationship between training sample size and model performance, we can determine the optimal amount of early-life data necessary for achieving accurate predictions and minimizing errors in PIML models.

As depicted in Fig. \ref{fig:Training_size}a, the performance of the PINN model shows a relatively consistent performance with slight improvements as more early-life experimental data points are included in the training process. This could be attributed to the utilization of late-stage simulation data in the training process, as well as the inclusion of constraints on the predicted half-cell model parameters within the loss function. In contrast, co-kriging exhibits a noticeable improvement in performance when transitioning from $n_{\mathrm{exp}}=90$ to $n_{\mathrm{exp}}=180$ training points. However, the performance gains become less prominent when further increasing the number of points from $n_{\mathrm{exp}}=180$ to $n_{\mathrm{exp}}=270$. This suggests that the additional training points beyond $n_{\mathrm{exp}}=180$ benefit less on enhancing the performance of the co-kriging model. 
We also observe similar trends for delta learning and data augmentation techniques, where utilizing $n_{\mathrm{exp}}=180$ characterization points from early-life experiments leads to decreased error for health parameters compared to using only 5 characterization points from each cell ($n_{\mathrm{exp}}=90$). Notably, the average results obtained with $n_{\mathrm{exp}}=180$ training points are comparable to those achieved when considering a greater number of training points. These findings align with the patterns observed in the PINN and co-kriging models, reinforcing the notion that an initial increase in training sample size can yield substantial improvements in model performance. This suggests that the additional information provided by $n_{\mathrm{exp}}=180$ early-life experimental data points (8 months of cycle aging data) is sufficient to capture key degradation patterns and reduce the prediction error for health parameters. However, adding more high-fidelity experimental data beyond a certain point does not significantly impact the model's performance. This can be attributed to the fact that the models have already been informed with the physics of late-life stages through the half-cell model. As a result, the incorporation of additional training points may not provide substantial new insights or improve the models' predictions.

The baseline model results are shown in Fig. \ref{fig:Training_size}b. When the number of experimental training points increased from 5 to 15 per cell, we observed a consistent trend of accuracy improvement. The neural network and elastic net baseline models both show significant improvement with this training size increase from $n_{\mathrm{exp}}=90$ to $n_{\mathrm{exp}}=180$ data points. However, further increasing to a total of $n_{\mathrm{exp}}=270$ experimental training points did not significantly reduce the error rates. In contrast, the baseline kriging model maintained consistent accuracy, regardless of the number of experimental training data points.

\subsection{Extrapolation Performance}
In this section, we delve into the extrapolation performance of the proposed PIML models and highlight the enhanced accuracy achieved by incorporating a physics-based half-cell model during the training process, specifically for the late-life stage. We conduct an analysis focusing on two novel approaches, namely PINN and co-kriging, for degradation diagnostics. To accomplish this, we compare the extrapolation performance of these methods to the purely data-driven counterpart models for a given test cell, considering two different conditions.

\begin{figure} 
\includegraphics[width=1\textwidth]{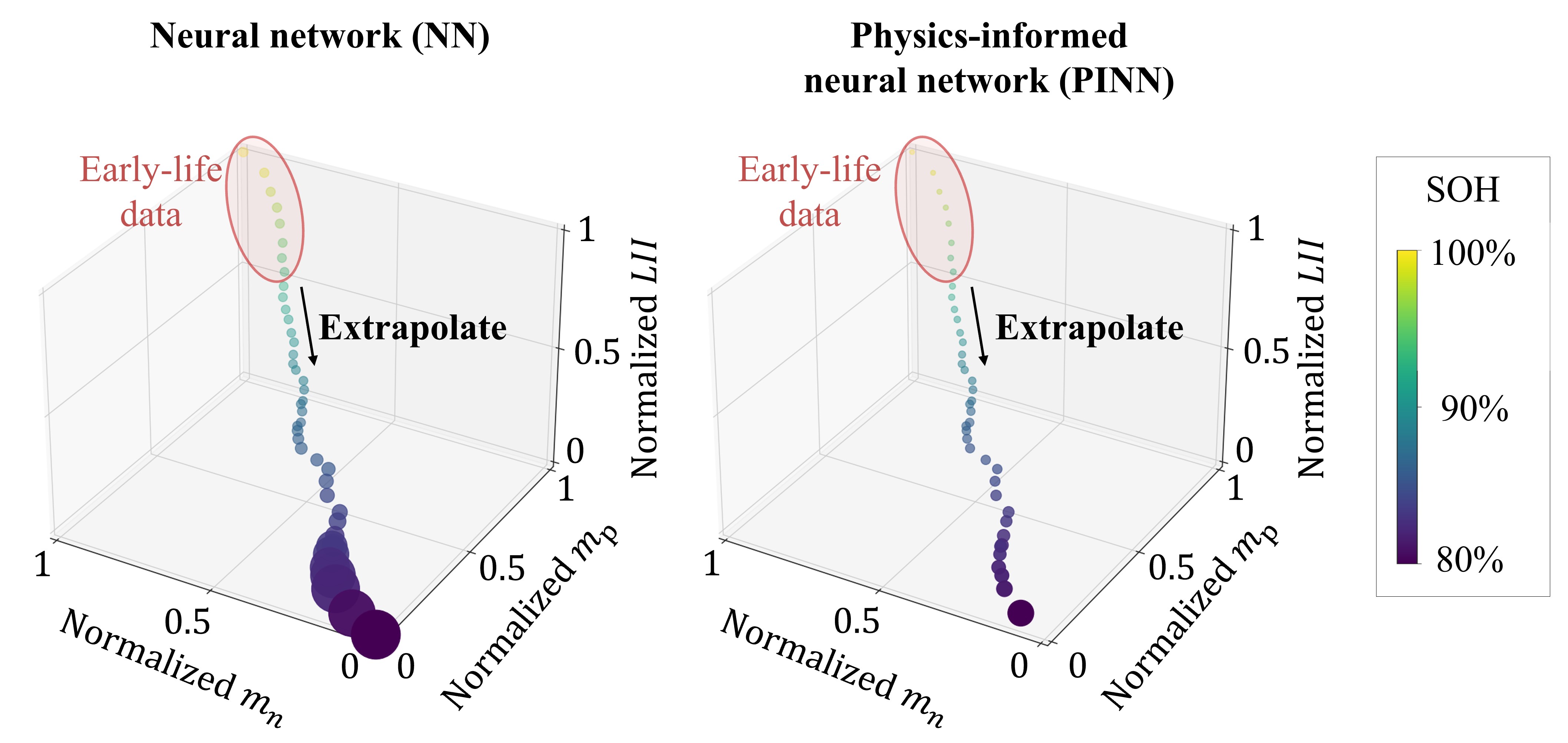}
\centering
\caption{Extrapolation error comparison for purely data-driven and physics-informed neural networks. The bubble size represents the RMSE computed based on the three degradation parameters. }
\label{fig:extrapolation_PINN}
\end{figure}

As mentioned in Sec. \ref{subsec:PINN}, we reinforce the underlying mapping between the $dQ/dV (V)$ curves and true health parameters into the PINN by adding the input data from the high-degradation region. Additionally, we constrain the {predicted half-cell model} parameters using a half-cell surrogate model. Furthermore, in the loss function, we perform peak difference minimization between the resulting $dQ/dV (V)$ curves obtained from the predicted half-cell model parameters and the experimental ones from the training set. To showcase the effect of incorporating these enhancements into the baseline model, we compare the extrapolation error of a specific test cell (G2C2) with a C/24 discharge rate and a temperature of $55^{\circ}$C. We analyze the extrapolation error in the degradation parameters space for both the physics-informed and regular neural networks. The results are visualized in Fig. \ref{fig:extrapolation_PINN}, where the bubble size represents the overall error, given by:

\begin{equation}
C_1 + C_2 \times RMSE([\hat{m}_\mathrm{p}, \hat{m}_\mathrm{n}, \hat{LII}]^\mathrm{T}, [m_\mathrm{p}, m_\mathrm{n}, LII]^\mathrm{T}) 
\end{equation}

Here, $C_1$ and $C_2$ are constants, and the root mean squared error (RMSE), denoted as $RMSE$, is computed based on the three degradation parameters: $m_{\mathrm{p}}$, $m_{\mathrm{n}}$, and $LII$. Based on the analysis, it is evident that as the degree of cell degradation increases, the extrapolation error of the regular neural network increases significantly compared to the physics-informed approach, particularly in the medium and late-life stages. This observation demonstrates the effectiveness of incorporating late-life stage degradation trends by augmenting the neural network with high degradation simulation data and imposing constraints to ensure consistency with the physics-based half-cell model during the training process. This integration enables the physics-informed network to capture and generalize the degradation patterns more accurately, leading to improved predictions even in the presence of severe degradation.

\begin{figure} [h]
\includegraphics[width=1\textwidth]{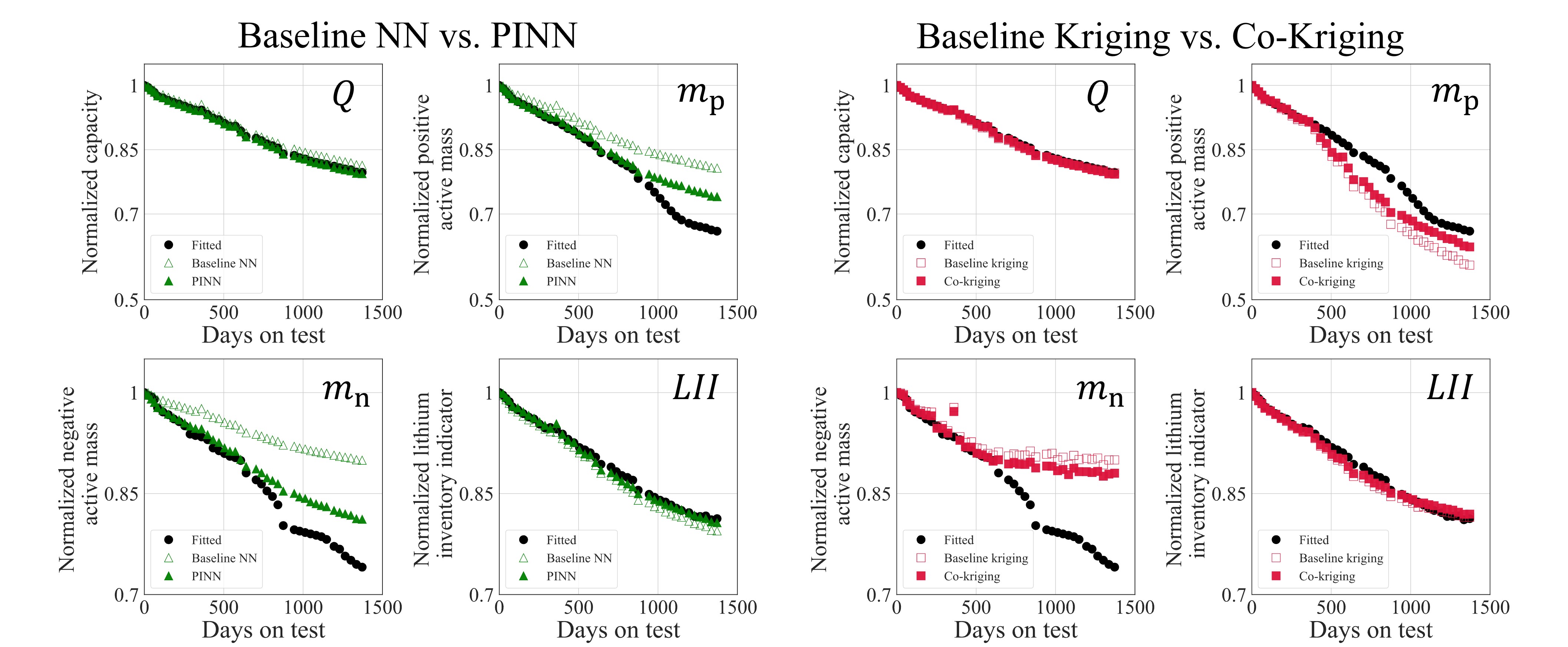}
\centering
\caption{Trajectory prediction comparison between purely data-driven and physics-informed machine learning models.}
\label{fig:PIML_vs_baseline_trajectory}
\end{figure}

To gain a better understanding of the extrapolation error, we visualize the predicted trajectory of each health parameter against its true trajectory obtained from the half-cell fitting process. The detailed results of trajectory predictions by the considered PIML methods for all cells can be found in the Appendix. Here, we focus on presenting the results for G3C3, a test cell with a higher discharge rate of C/3 and a temperature of $37 ^{\circ}$C. Fig. \ref{fig:PIML_vs_baseline_trajectory} illustrates the predicted trajectory of the four health parameters using both the PINN and co-kriging models compared to their baseline data-driven models. The predicted trajectory of the health parameters by these models exhibits significant improvement in tracking the true fitted values, particularly in the case of the positive and negative active mass parameters ($m_{\mathrm{p}}$ and $m_{\mathrm{n}}$), when compared to the baseline models. While the trajectory prediction for the parameters $Q$ and $LII$ also shows improvement, it is relatively less pronounced compared to the other two parameters. This indicates that incorporating trends from the high-degradation region and constraining machine learning models with our physics-based half-cell model is likely to improve the extrapolation of diagnosed degradation trends in the late-life stage, especially when dealing with more intricate patterns, such as those observed in $m_{\mathrm{p}}$ and $m_{\mathrm{n}}$.

\subsection{Sensitivity Analysis}
In this section, we perform an analysis on the sensitivity of two novel approaches, PINN and co-kriging, proposed in this study for degradation diagnostics. We aim to examine the impact of modifying their structure or parameters on the diagnostics error rate specifically during the late-life stage. For the PINN model, we focus on adjusting the weights assigned to each of the three loss terms incorporated in the physics-informed loss function. As for the co-kriging model, we assess the consequences of altering the kernel functions utilized for estimating and correcting within each kriging model.

\subsubsection{PINN} 
\subsubsubsection{Effect of Loss Terms} \mbox{}\\ \\
As described in Sec. \ref{subsec:PINN}, our implementation of the PINN incorporates three distinct loss terms. The first term, denoted as ($\mathcal{L}_1$), quantifies the disparity between the network's predictions of the half-cell model parameters and the actual fitted data. Additionally, we utilize two physics-informed loss terms. The second term ($\mathcal{L}_2$) arises from inputting the network's predictions into the half-cell model and ensuring that the obtained health parameters conform to the model's constraints. The third term ($\mathcal{L}_3$) considers the constraints on the peak positions of $dQ/dV (V)$ curves resulting from the half-cell model. These loss terms collectively contribute to the comprehensive training of our PINN model.

To evaluate the sensitivity of the PINN model's performance to each of these terms, we assess the ratio of their weights ($w_i$) in computing the total loss relative to the summation of all weights ($r_i$). This can be expressed as:

\begin{equation}
r_i = \frac{w_i}{\sum_{k=1}^{3} w_k} , \quad (i=1,2,3)
\end{equation}

\begin{figure} [h]
\includegraphics[width=1\textwidth]{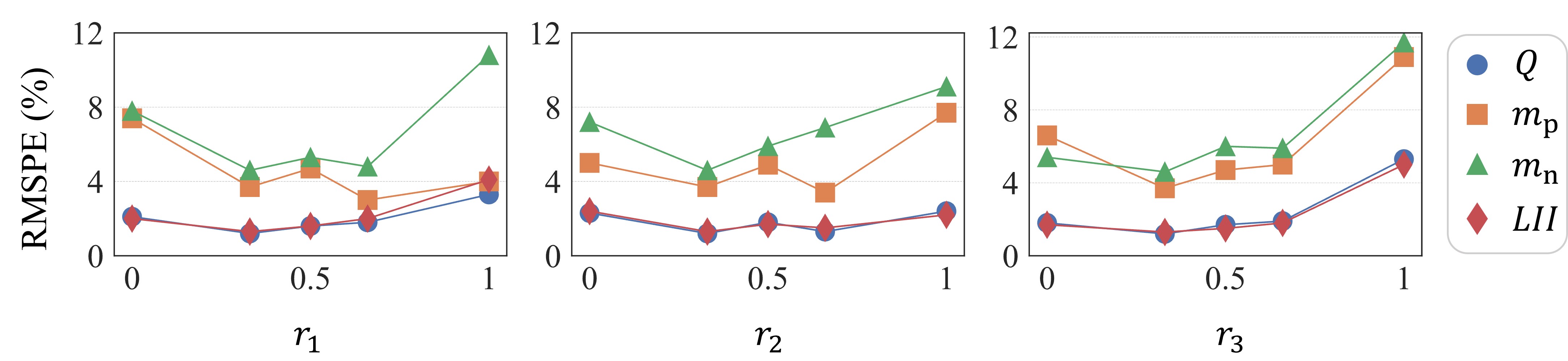}
\centering
\caption{Sensitivity of prediction error rate of PINN model to each loss term.}
\label{fig:Sensitivity_PINN}
\end{figure}

This ratio provides insight into the contribution and importance of each loss term in the overall training process of the PINN model. Initially, to ensure uniform scaling and equal impact of each loss term on the total loss, we assume a ratio of $r_i = 1/3 \approx 0.33$ for each of the weights. In Fig. \ref{fig:Sensitivity_PINN}, we plot the average prediction error rates for different values of this ratio corresponding to each loss term.

Five different ratios ranging from 0 to 1 are examined, as shown in Fig. \ref{fig:Sensitivity_PINN}. For each value of each ratio ($r_i$) considered, we modify one weight ($w_{i}$) while keeping the other two weights constant, contributing equally to the total loss. In cases where each loss term is considered alone ($r_i = 1$ for a specific loss term), the error rates tend to be higher. Notably, the first point in each subplot corresponds to the scenario where the specified loss term is not considered at all. The second point represents an equal weight assignment ($r_i = 1/3 \approx 0.33$) to all loss terms in the loss function. The third and fourth points correspond to $r_i = 1/2 = 0.5$ and $r_i = 2/3 \approx 0.66$, respectively. Remarkably, combining the three loss terms (represented by the three middle points in the subplots) results in lower overall error rates. For diagnosing $Q$, $m_{\mathrm{n}}$, and $LII$, assigning equal weights to the loss terms yields the minimum error rates. However, in the case of $m_{\mathrm{p}}$ diagnosis, increasing the ratio of the first and second loss terms between 0.5 and 1 can lead to more accurate results. These findings highlight the sensitivity of the PINN model's performance to the selection and weighting of different loss terms, emphasizing the importance of careful consideration when designing the loss function for specific degradation diagnostics tasks.

\begin{figure} [h]
\includegraphics[width=0.85\textwidth]{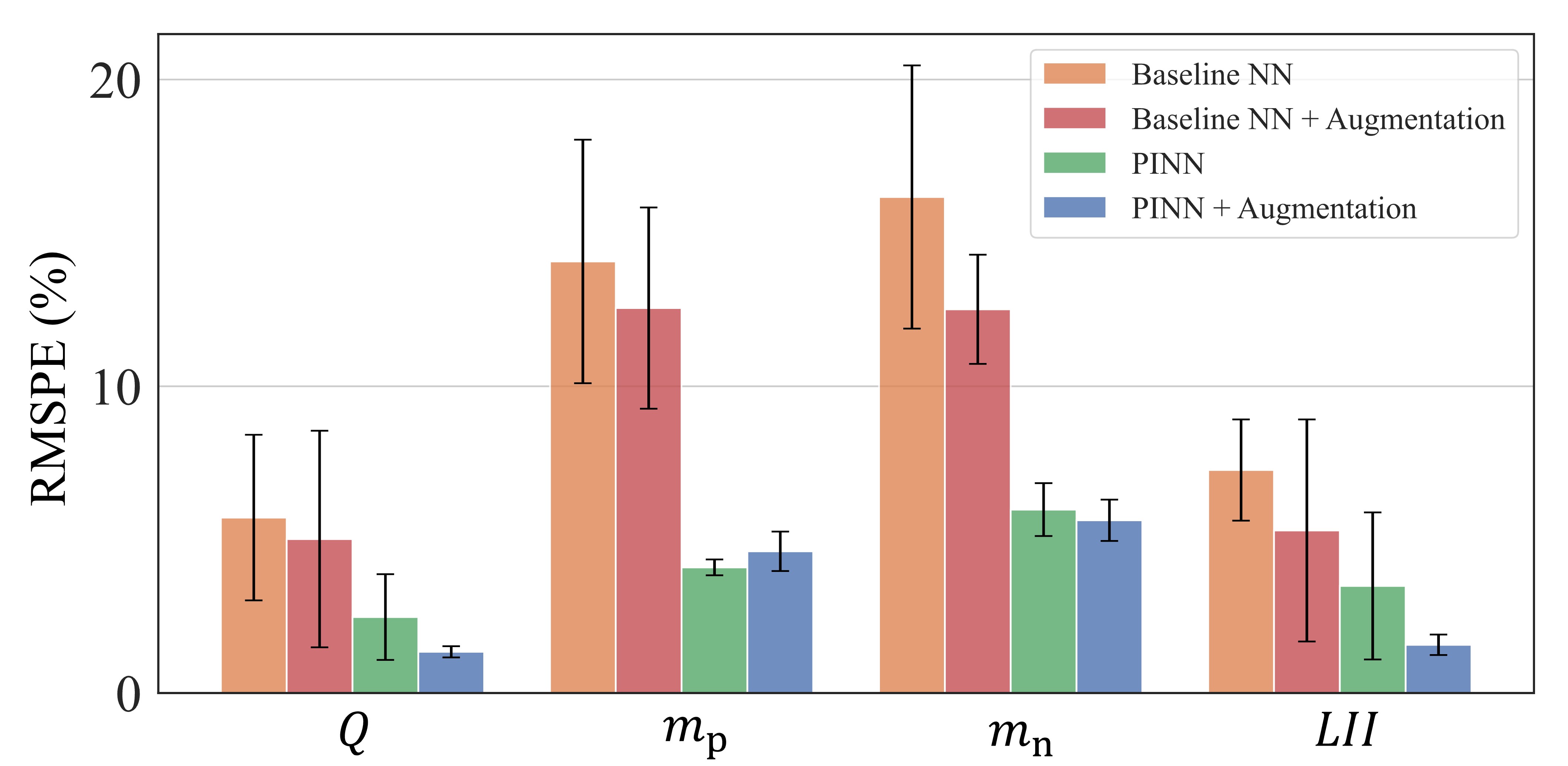}
\centering
\caption{Sensitivity of the PINN model to augmentation and the physics-informed loss function and sensitivity of the baseline NN model to augmentation.}
\label{fig:Sensitivity_PINN_2}
\end{figure}

\subsubsubsection{Effect of Loss Function vs. Augmentation} \mbox{}\\ \\
We conduct a sensitivity analysis to understand the specific impacts of two factors in enhancing the performance of the PINN model; the customized loss function and the augmentation of the training dataset. As mentioned in Sec. \ref{subsec:PINN}, the augmentation of the training dataset for the PINN model aimed to maintain consistency with other PIML approaches, ensuring a fair comparison. Additionally, this augmentation was intended to inform the PINN model about degradation trends in the late-life stage of the cells. While the simulation training data is generated from the same physical model, the primary enhancement in prediction accuracy stems from incorporating a customized loss function. To better demonstrate the contribution of each physics-informed technique, both the baseline neural network model and the PINN model were trained on only early-life experimental data, as well as on the augmented dataset that included the highest 20\% degradation from the simulation data, consistent with the settings for PINN model training. Ten repeated training runs were conducted under each training setup, and the RMSPE results averaged over these two runs are presented in Fig. \ref{fig:Sensitivity_PINN_2}. 

Two observations can be made from this figure. First, training the baseline model with the augmented dataset showed only marginal accuracy improvement in predicting health parameters, suggesting that the augmentation alone did not significantly enhance accuracy (Baseline NN vs. Baseline NN + Augmentation). Second, utilizing only early-life experimental data to train a PINN model (i.e., excluding the physics-based simulation data from the training set) resulted in only marginal increases in error rates (PINN vs. PINN + Augmentation).

These comparison results suggest that augmenting a training dataset with simulation data from the half-cell model and incorporating this model in the loss function do not yield the same effect in improving the ability of a neural network model to extrapolate. We can say, at least empirically from Fig. \ref{fig:Sensitivity_PINN_2}, that improved extrapolation capability can be mainly attributed to customizing the loss function with known physics. 

\subsubsection{Co-Kriging}
\begin{figure} [h]
\includegraphics[width=1\textwidth]{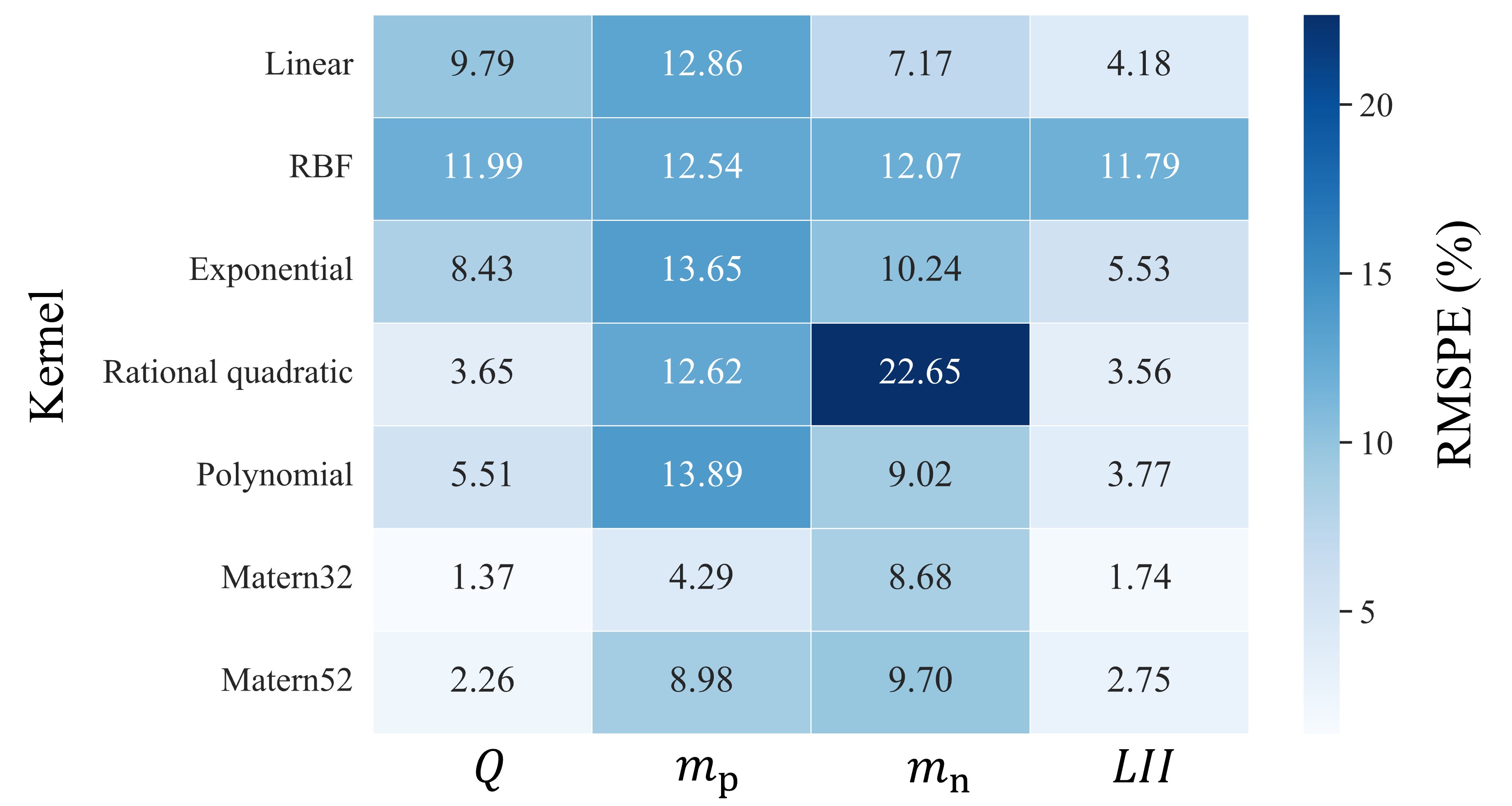}
\centering
\caption{Sensitivity of prediction error rate of co-kriging model to different kernel functions.}
\label{fig:Sensitivity_CK}
\end{figure}

As outlined in Sec. \ref{subsec:Co-kriging}, the co-kriging approach allows for the consideration of various types of kernel functions in each kriging model, depending on prior knowledge and understanding of the problem. The kernel function plays a crucial role in determining the correlation between data points and influences the accuracy of interpolation and extrapolation at unobserved locations. Different kernel functions possess distinct properties and assumptions. Here, we examine the impact of selecting different kernel functions on the prediction accuracy of the co-kriging approach in the context of degradation diagnostics. By evaluating the use of various kernels, we gain insights into the performance and effectiveness of the co-kriging method. The results of employing different kernel functions are presented and summarized in Fig. \ref{fig:Sensitivity_CK}. This analysis allows us to assess the influence of kernel selection on the overall predictive capabilities of the co-kriging approach, aiding in the determination of the most appropriate kernel function for our specific degradation diagnostics problem. A detailed description of the tested kernel functions is provided in the \ref{Appendix}.


As shown in Fig. \ref{fig:Sensitivity_CK}, the co-kriging approach using the Matern32 kernel function for each kriging model exhibits the lowest average error rates across all predicted health parameters. The Matern32 kernel is a well-suited option for certain data types due to its moderate smoothness and flexibility. It effectively models functions with a moderate level of smoothness while accommodating variability and irregular patterns commonly found in real-world data.

Belonging to the Matern family of kernels, the Matern32 kernel offers a range of smoothness options. With a smoothness parameter of 3/2, it strikes a balance between the extremely smooth RBF kernel (with a smoothness parameter of $\infty$) and the more oscillatory Matern52 kernel (with a smoothness parameter of 5/2). This moderate level of smoothness allows the model to capture variations and fluctuations in the data without overly constraining it. Consequently, the Matern32 kernel provides the necessary flexibility to accommodate different patterns and irregularities present in the data, making it a suitable choice for achieving accurate predictions. However, it is important to note that the polynomial kernel outperforms the Matern32 kernel in predicting $m_{\mathrm{n}}$. The Matern32 kernel's moderate smoothness assumption may not be ideal for data exhibiting stronger non-linearities, as observed in the degradation patterns of the negative electrode. On the other hand, the polynomial kernel offers the flexibility to represent more complex functions with higher-order polynomials and does not impose any smoothness constraint. This characteristic makes it well-suited for capturing more intricate variations and non-linear relationships in the data, making it a better choice for accurately modeling the $m_{\mathrm{n}}$ parameter.

Overall, choosing the right kernel for kriging and co-kriging depends on understanding the data's smoothness, trends, and noise levels. For smoother data, the RBF or Matern32 kernels are suitable, while non-linear patterns are better captured by the Matern52 or polynomial kernels. Evaluating different kernels, considering data complexity, incorporating domain knowledge, and performing cross-validation are vital for achieving accurate predictions while maintaining model interpretability and alignment with the underlying phenomenon.

\section{Discussion}
\subsection{Qualitative Analysis of PIML Techniques}
Selecting the appropriate PIML method for diagnosing battery degradation and general predictive purposes depends on several critical factors. These include the availability of data, the complexity of the specified process, and the desired level of accuracy and interpretability of the obtained results \cite{meng2022physics, karniadakis2021physics, xu2022physics}. Each PIML approach discussed in this study has its strengths and limitations, arising from their distinct approaches to integrating physics-based knowledge and data-driven techniques. A qualitative comparison of PIML methods is presented in Table \ref{table:Qual_compare}. In the subsequent paragraphs, we delve into the specific factors that have been evaluated in Table \ref{table:Qual_compare} for each PIML method.

\begin{table}[ht]
\caption{A qualitative comparison of state-of-the-art PIML approaches covered in this study}
\label{table:Qual_compare}
\centering
\adjustbox{max width=\textwidth}{
\small
\begin{tabular}{lcccc}
\toprule
\textbf{Quantity of Interest} & \textbf{PINN} & \textbf{Co-Kriging} & \textbf{Delta Learning with Enet} & \textbf{Data Augmentation} \\
\hline
Model Flexibility                           & High & Medium & Medium & Medium \\
Data Requirements                           & Low & Medium-low  & Medium-low  & Medium \\
Ease of Implementation                      & Low & Medium & High & High \\
Computational Cost                          & Low & Low & Low & Low \\
Generalization Capability                   & High & Medium & Medium & Medium \\
Interpretability                            & High & Medium-low & Medium-low & Low \\
Prediction Accuracy                         & High & High & High & High \\
Uncertainty Quantification Capability       & Medium-low & High & Low & Low \\
Scalability to Large Applications           & High & High & High & High \\
Applicability to Different Problem Types    & Medium-low & Medium-low & Medium & High \\

\bottomrule
\end{tabular}
}
\end{table}

\subsubsection{Model Flexibility}
Model flexibility is a key consideration in PIML methods, and it refers to the capability of each method to adapt to various types of physical models. PINNs, in this context, demonstrate a high degree of flexibility due to their ability to directly incorporate complex physics-based mappings and equations, including PDEs, into their loss function during the training process. A multitude of extensions for PINNs have been explored in the literature, encompassing areas such as energy conservation laws \cite{jagtap2020conservative}, finite element method \cite{xue2020amortized} and stochastic and fractional PDEs for multi-scale problems \cite{yang2020physics, pang2019fpinns}. This distinctive capability empowers PINNs to adeptly capture intricate physical relationships, thus reducing the reliance on solely simulated physics-based data. Conversely, other PIML techniques considered here, may require synthetic simulation data as an integral part of their training process. This dependence on simulated data can potentially impose limitations, particularly when confronted with highly complex equations characteristic of inverse and ill-posed problems \cite{karniadakis2021physics}. 
\subsubsection{Data Requirements}
The data requirements for each PIML method are vital considerations, including factors such as the quantity of data needed, data quality, and the availability of high-fidelity data for calibration or validation. These requirements vary among the methods, necessitating a comprehensive understanding of their effective implementation in practical applications. For PINNs, the amount of data required depends on the depth of the neural network used. Shallow neural networks, as employed in this study, may not demand an excessively large dataset for training. However, a notable strength of PINNs is their ability to achieve high accuracies even with limited amounts of data \cite{raissi2019physics}. This characteristic makes PINNs particularly valuable in situations where data availability is limited, providing a significant advantage over other methods. In contrast, delta learning, as a hybrid multi-fidelity model, relies heavily on high-fidelity data, which might be scarce compared to the more easily accessible low-fidelity data points \cite{aykol2021perspective}. To address this challenge, the delta learning approach introduces a corrector model to refine estimations based on low-fidelity data, making it more feasible to achieve accurate results even with limited high-fidelity data. Careful selection and filtering of the high-fidelity data are crucial in ensuring its effectiveness in enhancing the overall model accuracy \cite{yang2019physics}. Data augmentation, on the other hand, relies on low-fidelity synthetic data generated from a physics-based model. The success of data augmentation largely hinges on the relevance and diversity of the synthetic data points used \cite{chandrasekaran2023gait, meng2022physics}. Therefore, meticulous consideration in generating synthetic data is vital to achieve meaningful and reliable results.
\subsubsection{Ease of Implementation}
Ease of implementation is another factor when selecting a PIML method for a specific problem \cite{meng2022physics, xu2022physics}. Some PIML approaches offer straightforward algorithms or readily available libraries, which significantly simplifies their implementation process. In this context, both delta learning and data augmentation can be easily applied using the readily available machine learning packages in Python and MATLAB \cite{aykol2021perspective}. However, PINNs present a relatively more challenging implementation compared to delta learning and data augmentation. Customizing the loss function in a way that incorporates the physics-based constraints on the network outputs and ensuring consistency for gradient computation requires careful attention. Furthermore, co-kriging demands more consideration towards hyperparameter tuning and kernel selection \cite{yang2019physics}, compared to delta learning (when using simpler models like the elastic net), and data augmentation.
\subsubsection{Computational Cost}
The main challenges of using PIML for online prediction are the cost and time required for training and inference processes \cite{xu2022physics}.  Computational cost is inherently intertwined with complexity of the physics-based model and the machine learning model \cite{aykol2021perspective}. In the context of the study, where online estimation applications are targeted, all the considered methods have been specifically designed and implemented with computational efficiency in mind.  
\subsubsection{Scalability to Large Applications}
Scalability assesses the ability of each method to scale with increasing data sizes and problem complexities. The ability to scale effectively is particularly important for real-world applications, where datasets can be vast, and the problems may involve intricate physical processes. The PIML methods demonstrated the capability to effectively handle larger dataset sizes and more complex problem scenarios within the context of this study, as well as for other problem domains \cite{kashinath2021physics}. However, when dealing with larger-sized problems, careful considerations and customizations are required to ensure optimal performance. Scaling to larger datasets and more complex problems may involve adjustments in computational resources, algorithmic optimizations, or fine-tuning of hyperparameters. It should be noted that among the considered methods here, data augmentation emerges as a relatively more straightforward approach to scale from small-scale problems to larger ones \cite{pombo2022benchmarking}. The inherent advantage of data augmentation lies in its ability to utilize expanded dataset, allowing for seamless extensions to larger datasets without significant constraints on data availability.
\subsubsection{Generalization Capability}
Generalization capability is another crucial factor that investigates how well each method can generalize to unseen data or different operating conditions while considering issues such as overfitting. In this regard, PINNs exhibit superior performance compared to other PIML methods. Unlike other methods that rely solely on generated physics-based data for training, PINNs can directly incorporate physics-based constraints into the loss function, enhancing their ability to generalize to unseen data and different operating conditions \cite{xu2022physics, drgovna2021physics}.In contrast, delta learning and data augmentation depend on specific data trends during training, leading to relatively lower generalization capabilities.   
\subsubsection{Interpretability}
Interpretability examines how well each method allows for interpretability and physical insights. Some methods provide explicit representations of the underlying physical processes, while others may have more black-box nature \cite{rudin2019stop}. PINNs excel in this regard, as the physics-based constraints are explicitly incorporated into the loss function, enabling clear mathematical descriptions of their influence on the model outputs \cite{xu2022physics}. On the other hand, data augmentation may lack the same level of interpretability since its reliance on synthetic data may not directly reveal the underlying relationships between inputs and outputs. 
\subsubsection{Prediction Accuracy}
Traditional machine learning methods often lack the ability to extrapolate, which can be solved by the PIML methods \cite{zhao2019physics}. Prediction accuracy evaluates the predictive performance of each PIML method on both training and test datasets, which, as discussed extensively in this study, all the PIML methods showed promising results. 
\subsubsection{Applicability to Different Problem Types}
Applicability to different problem types assesses whether each method is well-suited to specific problem types or domains. In this regard, PINNs can be more challenging to apply in some cases, as the explicit definition of physics-based mapping functions between inputs and outputs may not be readily available \cite{karniadakis2021physics}. However, PINNs become a viable option when generating training data from physics-based models is more difficult than incorporating equations into the loss function and training process. Similarly, co-kriging faces challenges when defining correlations between different fidelity levels of data is not straightforward \cite{brevault2020overview}. Conversely, delta learning, in general, exhibits adaptability to different problem structures by selecting appropriate regression models, making it applicable to various scenarios. On the other hand, data augmentation stands out as the most versatile approach across various fields and applications. Its ability to use synthetic simulation data from physics-based models for training purposes makes it a straightforward and effective choice \cite{meng2022physics}.
\subsubsection{Uncertainty Quantification Capability}
Lastly, uncertainty quantification examines whether each method provides a means to quantify uncertainties in predictions. In the case of PINNs, as they are not inherently probabilistic, methods like ensembles \cite{lakshminarayanan2017simple}, quantile regression \cite{cannon2011quantile}, and Bayesian neural networks \cite{mullachery2018bayesian,nemani2023uncertainty} should be considered to quantify uncertainty.  Each of these techniques offers a unique perspective on capturing and expressing uncertainties inherent in predictive modeling, particularly in scenarios involving short-term data. Ensembles involve training multiple models independently and aggregating their predictions. Quantile regression is a technique that enables the estimation of different quantiles of the predictive distribution. In the context of PINNs, quantile regression allows for modeling a range of possible outcomes and their associated uncertainties. Bayesian neural networks present another avenue for incorporating uncertainty into PINNs. Unlike conventional neural networks, Bayesian neural networks treat model parameters as probability distributions. This probabilistic approach allows for the modeling of parameter uncertainty, offering a more nuanced representation of the model's predictions.
On the other hand, co-kriging is a powerful approach to uncertainty quantification. It naturally accounts for uncertainty by modeling correlations and providing prediction variances \cite{bilionis2012multi}. Co-kriging explicitly models the covariance structure between different variables, capturing the correlation between their variations. The incorporation of a well-defined covariance structure enables co-kriging to effectively propagate uncertainties across variables \cite{nemani2023uncertainty}.
For delta learning and data augmentation with elastic net model, while they may not inherently provide uncertainty estimates, it is possible to introduce uncertainty quantification methods such as Bayesian \cite{crandell2011bayesian} and bootstrap methods \cite{endo2015confidence} to enhance the prediction reliability and quantify uncertainties. 

In summary, making informed decisions based on these factors and considerations is required to effectively apply PIML methods in different scenarios. In addition, ongoing research and advancements in various PIML methods continue to expand their applicability and improve their performance, making it important to stay updated with the latest developments in the field. 

\subsection{Method Selection Guidelines}
PIML techniques are primarily applied to address and solve physics or engineering problems in situations where acquiring data and formulating tasks can be particularly challenging for researchers who lack domain knowledge and experience. Moreover, current research heavily depends on domain-specific datasets, complicating the identification of best practices for specific problems. Here, we will succinctly summarize the conclusions drawn from our comprehensive study and the reviewed literature, offering insights into when to choose each PIML technique.
\subsubsection{PINN}
PINNs excel in situations where the underlying physics of the problem can be represented by a set of differentiable mapping functions. Although PINNs can achieve high accuracy with a limited amount of high-fidelity data, it is important to note that implementing this method requires more intricacy compared to other PIML approaches in terms of both mathematics and programming resources \cite{xu2022physics}. Creating a loss function that incorporates physics-informed terms with a closely matched convergence rate, while also ensuring that the neural network's computational graph is interconnected to facilitate gradient calculations for outputs with respect to inputs, can be a challenging task \cite{wang2020understanding}. Therefore, careful consideration and expertise are needed to effectively employ PINNs for a given problem. However, once implemented, PINNs tend to generalize well to new, unseen data and are less prone to overfitting. In the specific case of degradation diagnostics for lithium-ion batteries, the prediction results of the PINN model for health parameters at a late-life stage show a significant improvement compared to the purely data-driven counterpart. When compared to other PIML approaches presented, the PINN methodology demonstrated superior accuracy and consistency in predicting more intricate degradation trends, specifically in the loss of positive and negative active mass ($m_{\mathrm{p}}$ and $m_{\mathrm{n}}$).

\subsubsection{Delta Learning}
Delta learning offers ease of implementation and flexibility by allowing the selection of any regression machine learning model as the estimator and corrector models. This adaptability makes it an attractive option, especially when the physical model is not explicitly known, but limited high-fidelity data is available through experiments. It's crucial to select and evaluate machine learning models carefully to ensure accuracy and reliability in the delta learning approach. To achieve optimal results, diverse sampling of high-fidelity points across the entire design space is important, particularly for accurate extrapolation.

\subsubsubsection{Delta Learning with Elastic Net Regression} \mbox{}\\ \\
For cases where predicted trends are not overly intricate, opting for simpler models like elastic net can be advantageous. Elastic net effectively captures underlying patterns and provides accurate predictions. This is particularly valuable when prioritizing ease of implementation and time efficiency. In the specific study's case of predicting capacity ($Q$) and lithium inventory indicator ($LII$), the elastic net model showed good performance. The choice of elastic net is also suitable due to the small size of the training data, which reduces the risk of overfitting associated with more complex models \cite{thelen2022integrating}.

\subsubsubsection{Delta Learning with Kriging (Co-Kriging)} \mbox{}\\ \\
Incorporating kriging (GPR) models introduces a powerful approach known as co-kriging, which is effective in modeling multi-fidelity data problems. Co-kriging leverages correlated fidelity data to fill in missing values and estimate uncertainties in predicted values. It facilitates the quantification of result reliability. The co-kriging method is particularly useful for predicting capacity ($Q$) and lithium inventory indicator ($LII$), as observed in this study. The capacity prediction using co-kriging exhibited a near-zero error rate, showcasing its high accuracy. Additionally, co-kriging showed a significant improvement in predicting the positive active mass parameter ($m_{\mathrm{p}}$).

\subsubsection{Data Augmentation}
In situations where the underlying physical model or correlations between different fidelity sources of data are not well understood or difficult to formulate explicitly, data augmentation offers a data-driven framework to improve model performance. By generating additional training samples, data augmentation caters to various cost considerations based on available resources. This approach is commonly used alongside deep learning models, enabling them to effectively capture complex non-linear relationships, as extensively documented in the literature. However, even with a simple elastic net model, data augmentation has demonstrated effectiveness, as evidenced in the case of this study. The method's ease of implementation and computational efficiency makes it particularly advantageous for rapid prototyping and large-scale applications.

\section{Conclusion}
In this study, we introduced two PIML approaches: PINN and co-kriging. These approaches were designed for diagnosing cell degradation without depending on late-life aging data. Through a comparative analysis, we evaluated these methods alongside delta learning and data augmentation. We used battery aging data from a long-term cycling experiment for this evaluation. The comparative results highlighted the superior performance of the PIML approaches compared to their purely data-driven baseline models. This performance improvement was especially notable in terms of extrapolation during the late-life stage of degradation. Additionally, we conducted a sensitivity analysis to gain deeper insights into implementing the two new approaches: PINN and co-kriging. We also engaged in a detailed discussion of the advantages and limitations of each method, grounded in the context of battery degradation diagnostics and existing literature. Furthermore, we provided guidance on when to choose a specific PIML method based on the insights gathered. This study contributes valuable knowledge to the realm of battery degradation diagnostics and PIML method selection.

\bibliographystyle{unsrtnat}
\bibliography{References}  

\clearpage
\section*{Appendix}
\label{Appendix}

\subsection*{A1. Degradation quantification using a half-cell model}

In a half-cell model, the pseudo-full-cell OCV curve can be constructed and mathematically expressed as follows, 
\begin{equation*}
\label{eqn:half_cell}
    V_{\mathrm{c}}(Q)|_{Q=Q_{\mathrm{c}}} = V_{\mathrm{p}}(q_{\mathrm{p}})|_{q_{\mathrm{p}}=\frac{Q-\delta_{\mathrm{p}}}{m_{\mathrm{p}}}} - V_{\mathrm{n}}(q_{\mathrm{n}})|_{q_{\mathrm{n}}=\frac{Q-\delta_{\mathrm{n}}}{m_{\mathrm{n}}}} 
\end{equation*}
where $V_{\mathrm{c}}(Q)$ is the pseudo-full-cell OCV curve, $Q_{\mathrm{c}}$ refers to the cell capacity at different states of charge, $V(q)$ is a half-cell curve with specific capacity $q$ $(\si{mAh/g})$, $m$ is the active mass $(\si{g})$, and $\delta$ is the half-cell curve slippage $(\si{mAh})$. The subscripts $\mathrm{p}$ and $\mathrm{n}$ correspond to the positive and negative electrodes (PE and NE). The two slippage parameters, $\delta_{\mathrm{p}}$ and $\delta_{\mathrm{n}}$, quantify the horizontal distance of the left endpoint of the positive and negative half-cell curves to $Q_{\mathrm{c}}=0$. 

Then, the differential voltage curve ($dV/dQ (Q)$) of a half-cell model can be derived as follows,
\begin{equation*}
\label{eqn:half_cell_diff}
    \frac{dV_{\mathrm{c}}}{dQ}(Q)|_{Q=Q_{\mathrm{c}}} = \frac{1}{m_{\mathrm{p}}}\frac{V_{\mathrm{p}}(q_{\mathrm{p}})}{dq_{\mathrm{p}}}|_{q_{\mathrm{p}}=\frac{Q-\delta_{\mathrm{p}}}{m_{\mathrm{p}}}} - \frac{1}{m_{\mathrm{n}}}\frac{dV_{\mathrm{n}}(q_{\mathrm{n}})}{dq_{\mathrm{n}}}|_{q_{\mathrm{n}}=\frac{Q-\delta_{\mathrm{n}}}{m_{\mathrm{n}}}} 
\end{equation*}

In this work, we follow a manual approach, presented in \citep{thelen2022integrating}, to fit a half-cell model to cells by adjusting four half-cell parameters ($m_{\mathrm{p}}$, $m_{\mathrm{n}}$, $\delta_{\mathrm{p}}$, and $\delta_{\mathrm{n}}$). This manual process focuses on: 1. aligning peaks from the reconstructed 
$dV/dQ (Q)$ curves to the peaks from experimental $dV/dQ (Q)$ curves; 2. matching the reconstructed OCV curves to experimental $QV$ curves, which considers both the endpoint locations and the overall shape. First, $m_{\mathrm{n}}$ and  $\delta_{\mathrm{n}}$ are adjusted to match the peaks on $dV/dQ (Q)$ curves since the negative electrodes contribute major peaks in the OCV. After tuning the half-cell curve of the negative electrode, $m_{\mathrm{p}}$ and  $\delta_{\mathrm{p}}$ are used to adjust peak magnitudes on $dV/dQ (Q)$ curves and line up the endpoints of $QV$ curves. This fitting process is repeated to fine-tune the fitting results to match the fitted curves as much as possible.

\subsection*{A2. Automatic Fitting vs. Manual Fitting}
\begin{figure} [h]
    \centering
    \includegraphics[width=0.75\textwidth]{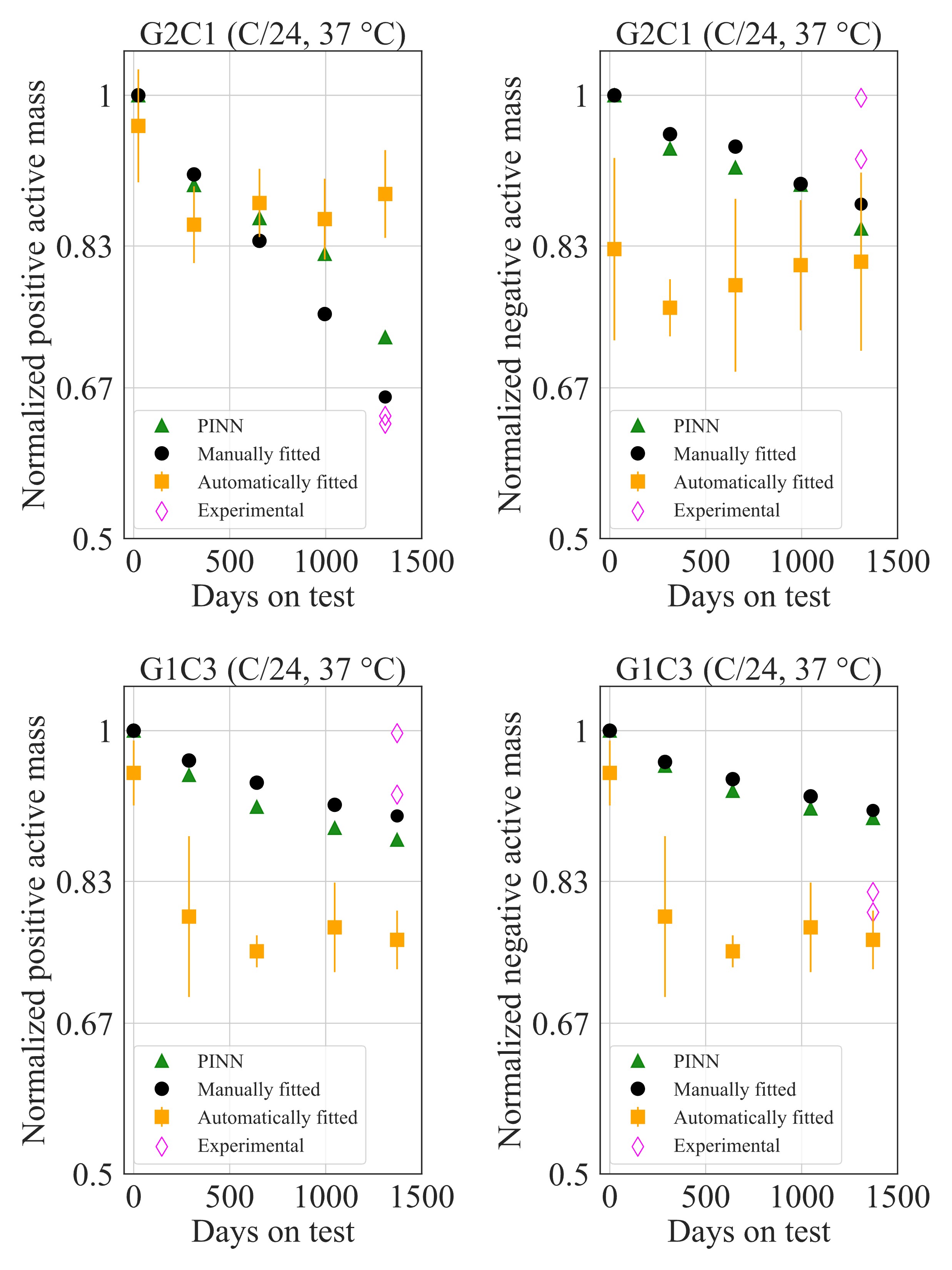}
    \caption{Comparison of active mass estimates by PINN, manual and automatic fitting, and experimental validation for two selected cells.}
    \label{fig:ManualvsAuto}
\end{figure}

Here, we conducted a comparative analysis between manual and automatic approaches for fitting the half-cell model to justify the necessity of the development of the PIML methods using the manual fitting approach. We compared the positive and negative positive mass estimates by manual fitting, as outlined in Sec. \ref{subsec:hc}, with half-cell parameter estimates by nonlinear optimization using the original half-cell potential data (i.e., automatic fitting). This comparison additionally included the PINN model, trained on parameter estimates by the manual fitting approach. It was performed on two cells destructively analyzed after certain aging periods as an experimental validation of positive and negative active mass estimations. 
Automatic fitting employed a bi-objective loss function to match the endpoints of the simulated and experimental voltage vs. capacity ($QV$) curves and improve the overall agreement between these two curves by minimizing an RMSE loss. Experimental validation initially involved recovering and treating two samples from each working electrode of the two selected cells. Then,  each sample was coupled with lithium metal to build coin half-cells. These coin half-cells were cycled at a slow rate to obtain their capacity. Finally, the full cell’s remaining positive and negative active mass was calculated based on the nominal specific capacity and area ratio \cite{lui2021physics}. 
The comparison results are depicted in Fig. \ref{fig:ManualvsAuto}, revealing certain limitations of nonlinear optimization (automatic fitting). 
\begin{itemize}
    \item First, the optimization problem for automatic fitting has multiple local minima, leading to run-to-run variability in optimal active mass parameters depending on the initial guess. We illustrated this variability by presenting the mean and error bars (spread) derived from five optimization runs, each starting at a different initial guess. 
	\item Second, manual fitting considers phase transitions by matching characteristic features (e.g., the position and height of a peak) of simulated and experimental $dQ/dV (V)$ curves. However, incorporating feature matching into nonlinear optimization could be challenging due to the need to automatically extract features and track the trend of how each half-cell model parameter evolves. 
	\item Third, even if such optimization were feasible, its implementation on a battery management system would be impractical due to the extensive tuning required. The use of an optimization strategy reliant on iterative tuning for feature matching is likely to surpass the time constraints imposed by real-time battery management requirements. Moreover, the challenge of tuning the system to generate reliable initial parameter estimates for diverse operating conditions introduces an additional layer of complexity.
\end{itemize}

In contrast, building machine learning models for degradation diagnostics reduces the time cost associated with manual fitting and, mostly importantly, automates health parameter estimation by taking a measured $dQ/dV (V)$ curve as the model input, running a forward pass of the model, and producing an estimate of the health parameters as the output. This automated process is particularly suitable for on-board battery management system applications where extensive tuning is infeasible.

Additionally, the comparison results in Fig. \ref{fig:ManualvsAuto} show that active mass estimates by the machine learning method (i.e., PINN) align well with those by manual fitting; the former offers the additional benefits of reduced time cost and automation. Both methods generally agreed better with experimental validation than automatic fitting. In conclusion, our machine learning approach strikes a practical balance between accuracy and feasibility for real-world applications.

\subsection*{A3. Kernel functions in GPR (kriging)}
\subsubsection*{Linear kernel}
The linear kernel models a linear relationship between input data points. Let $\mathbf{x}$ and $\mathbf{x'}$ be two input feature vectors. The linear kernel function, denoted as $k_{\mathrm{linear}}(\mathbf{x}, \mathbf{x'})$, is defined as the dot product of the two feature vectors:
\[ k_{\mathrm{linear}}(\mathbf{x}, \mathbf{x'}) = \mathbf{x}^\mathrm{T} \mathbf{x'} \]

\subsubsection*{Radial basis function (RBF) kernel}
The RBF kernel, also known as the squared exponential kernel, captures smooth, non-linear relationships between data points.It measures the similarity between two data points based on the distance between them in the feature space. It assigns higher similarity (covariance) to data points that are closer to each other and lower similarity to data points that are farther apart. The RBF kernel is commonly used when the assumption of smoothness and stationarity is reasonable in the data.
\[ k_{\mathrm{RBF}}(\mathbf{x}, \mathbf{x'}) = \sigma^2 \exp \left( -\frac{\lVert \mathbf{x} - \mathbf{x'} \rVert^2}{2l^2} \right) \]
where $\sigma$ is the kernel bandwidth or length scale parameter. It determines the "width" of the kernel and controls the smoothness of the resulting function.

\subsubsection*{Exponential kernel}
The exponential kernel is similar to the RBF kernel and also captures smooth, non-linear relationships. However, it tends to have a faster decay, making it more suitable for cases where data points are expected to be less correlated as the distance between them increases.
\[ k_{\mathrm{exp}}(\mathbf{x}, \mathbf{x'}) = \sigma^2 \exp \left( -\frac{\lVert \mathbf{x} - \mathbf{x'} \rVert}{l} \right) \]

\subsubsection*{Rational quadratic kernel}
The rational quadratic (RQ) kernel is a smooth non-linear kernel with a tunable parameter $\alpha$. It can capture a wide range of smoothness levels in the data and can be useful when the underlying relationship is not strictly linear or purely smooth.
\[ k_{\mathrm{RQ}}(\mathbf{x}, \mathbf{x'}) = \sigma^2 \left( 1 + \frac{\lVert \mathbf{x} - \mathbf{x'} \rVert^2}{2\alpha l^2} \right)^{-\alpha} \]
The value of $\alpha$ controls the balance between the linear and non-linear components of the kernel. When $\alpha$ is large, the RQ kernel behaves more like the RBF kernel, emphasizing non-linear patterns. On the other hand, when $\alpha$ is close to zero, the RQ kernel behaves more like the linear kernel, capturing linear relationships in the data.

\subsubsection*{Polynomial kernel}
The polynomial kernel models polynomial relationships between data points. It can be useful when the relationship between the inputs is known to be polynomial in nature. It is a generalization of the linear kernel and can capture various degrees of polynomial relationships between data points.
\[ k_{\mathrm{poly}}(\mathbf{x}, \mathbf{x'}) = (\mathbf{x}^\mathrm{T} \mathbf{x'} + c)^d \]
where $d$ is the degree of the polynomial, and $c$ is an optional constant term that can be included to control the bias of the polynomial expansion.

\subsubsection*{Matérn 3/2 kernel}

The Matern kernel family in GPR modeling includes two specific variants known as Matern32 and Matern52 kernels. Both kernels are characterized by their ability to model different levels of smoothness in the data, making them useful in various applications. The Matérn 3/2 kernel is used to model data with sharp changes and smoothness. It is commonly used when the data is expected to have abrupt variations, but some level of smoothness is also present, making it suitable for modeling real-world data with moderate roughness. It is also computationally efficient and suitable for datasets with a moderate number of data points.

\[ k_{\mathrm{Mat32}}(\mathbf{x}, \mathbf{x'}) = \sigma^2 \left(1 + \frac{\sqrt{3} \lVert \mathbf{x} - \mathbf{x'} \rVert}{l} \right) \exp \left( -\frac{\sqrt{3} \lVert \mathbf{x} - \mathbf{x'} \rVert}{l} \right) \]
where $\rho$ is the length scale parameter.

\subsubsection*{Matérn 5/2 kernel}
The Matern52 kernel is another member of the Matern family, with a smoothness parameter of $\nu = 5/2$ (differentiable twice). It is smoother than the Matern32 kernel and can effectively model data with higher roughness or irregularity.
\[ k_{\mathrm{Mat52}}(\mathbf{x}, \mathbf{x'}) = \sigma^2 \left(1 + \frac{\sqrt{5} \lVert \mathbf{x} - \mathbf{x'} \rVert}{l} + \frac{5\lVert \mathbf{x} - \mathbf{x'} \rVert^2}{3 l^2} \right) \exp \left( -\frac{\sqrt{5} \lVert \mathbf{x} - \mathbf{x'} \rVert}{l} \right) \]
where $\rho$ is the length scale parameter.

\subsection*{A4. Trajectory prediction results}
Here, we present the trajectory prediction results for all the test cells across six groups. The plots display the predicted health parameters generated by each of the four PIML methods considered, alongside the fitted true values.
\begin{figure} [h]
\includegraphics[width=0.81\textwidth]{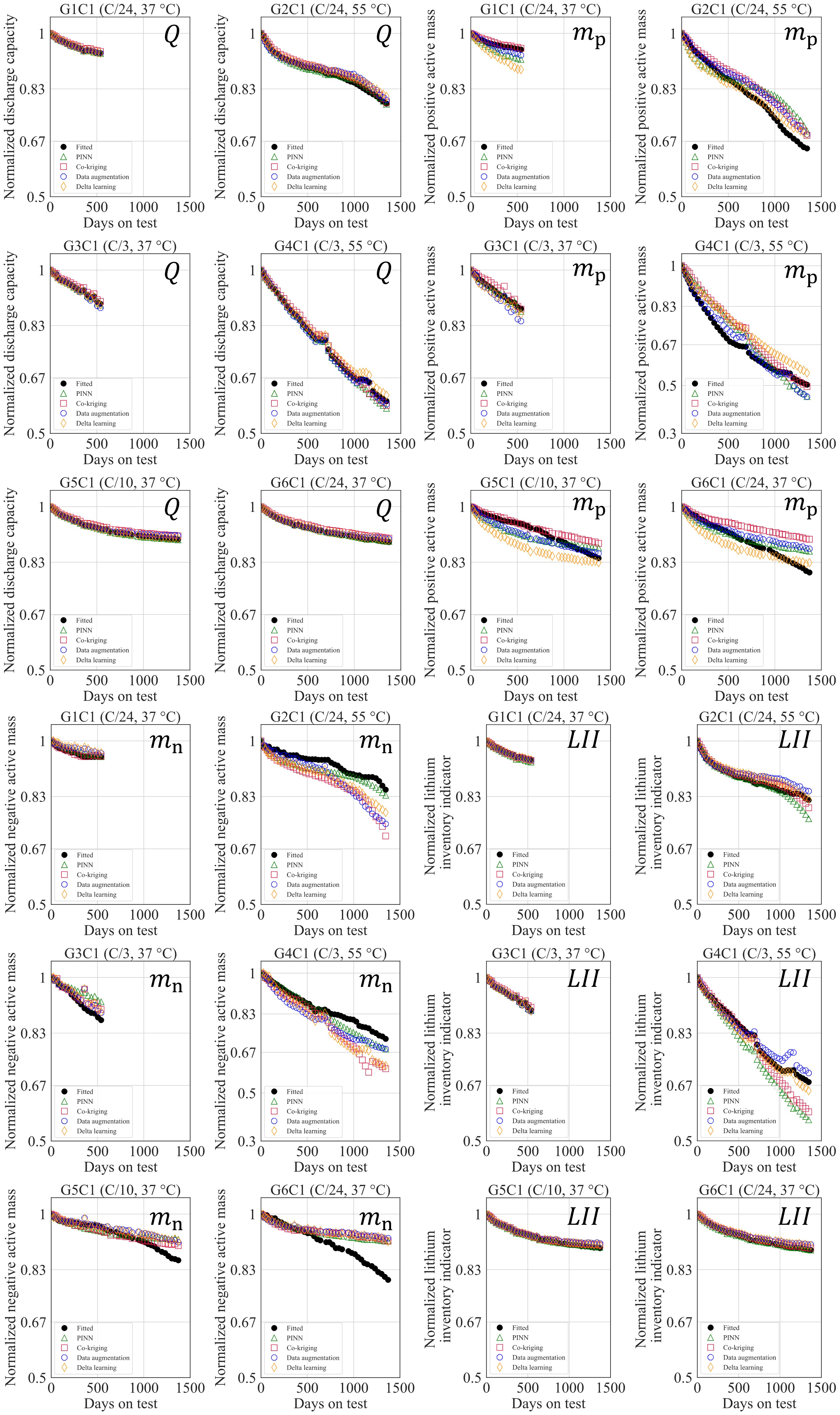}
\centering
\caption{Trajectory prediction results for the first fold (first cell of each group).}
\label{fig:Trajectory_1}
\end{figure}

\begin{figure} [h]
\includegraphics[width=0.81\textwidth]{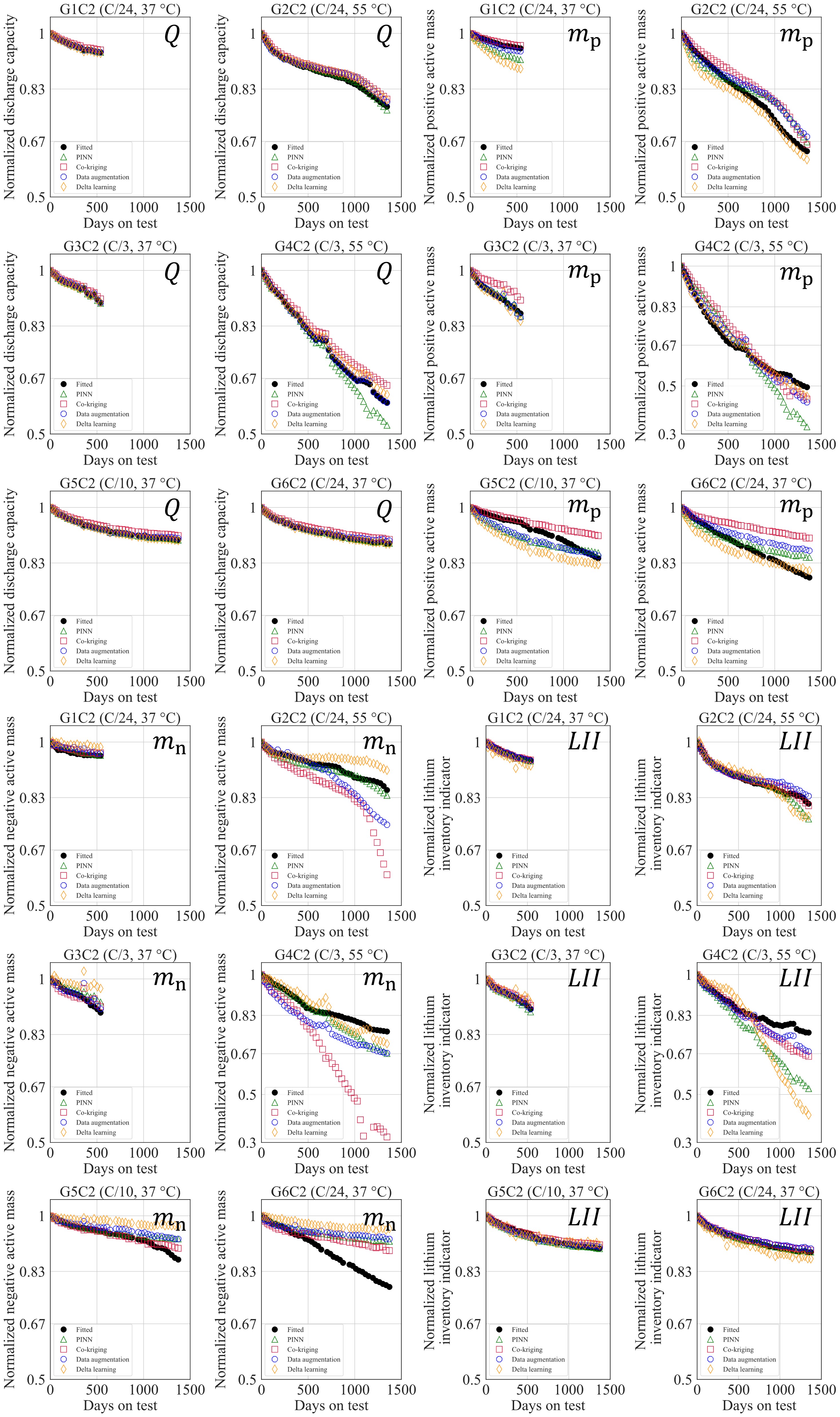}
\centering
\caption{Trajectory prediction results for the second fold (second cell of each group).}
\label{fig:Trajectory_2}
\end{figure}

\begin{figure} [h]
\includegraphics[width=0.81\textwidth]{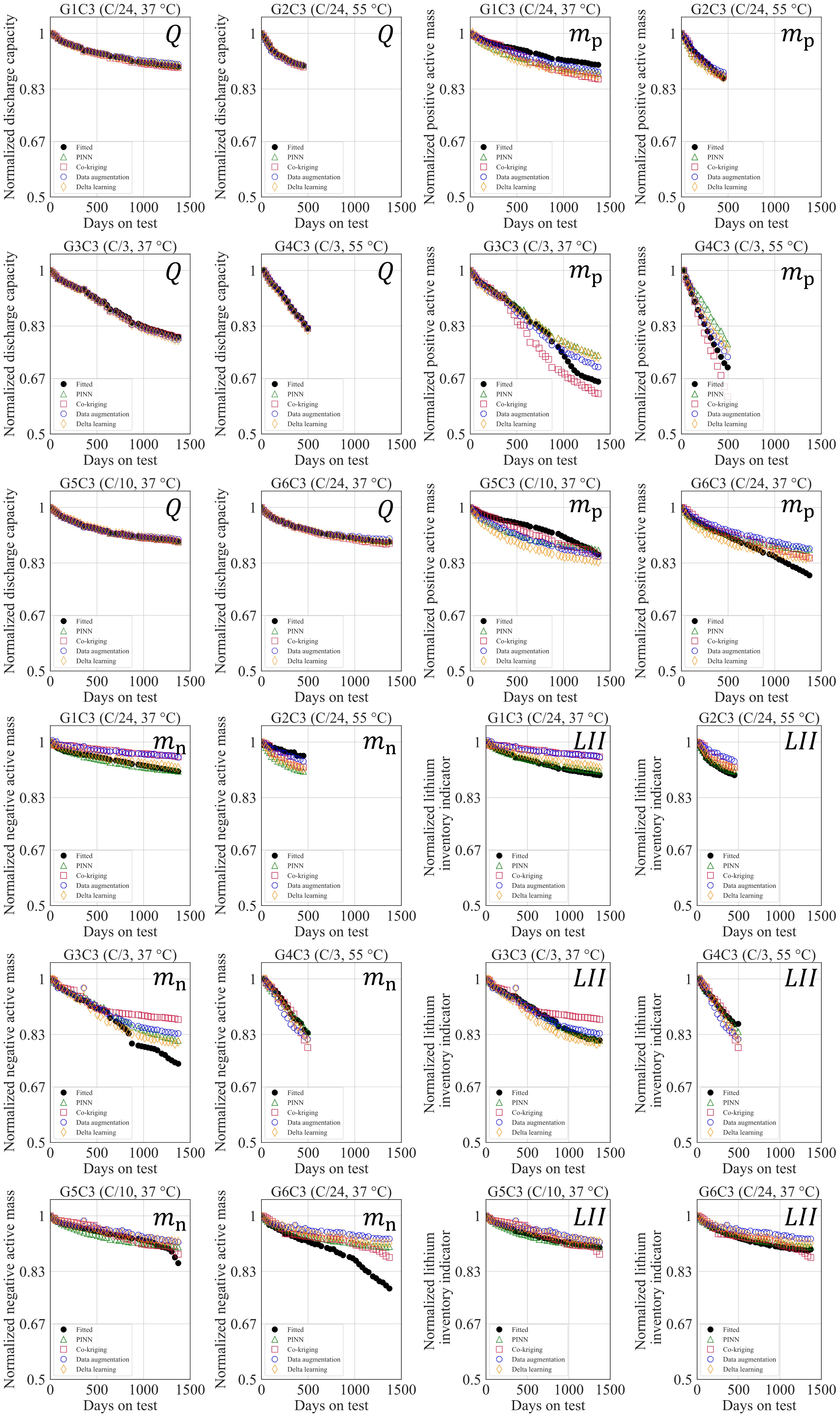}
\centering
\caption{Trajectory prediction results for the third fold (third cell of each group).}
\label{fig:Trajectory_3}
\end{figure}

\begin{figure} [h]
\includegraphics[width=0.81\textwidth]{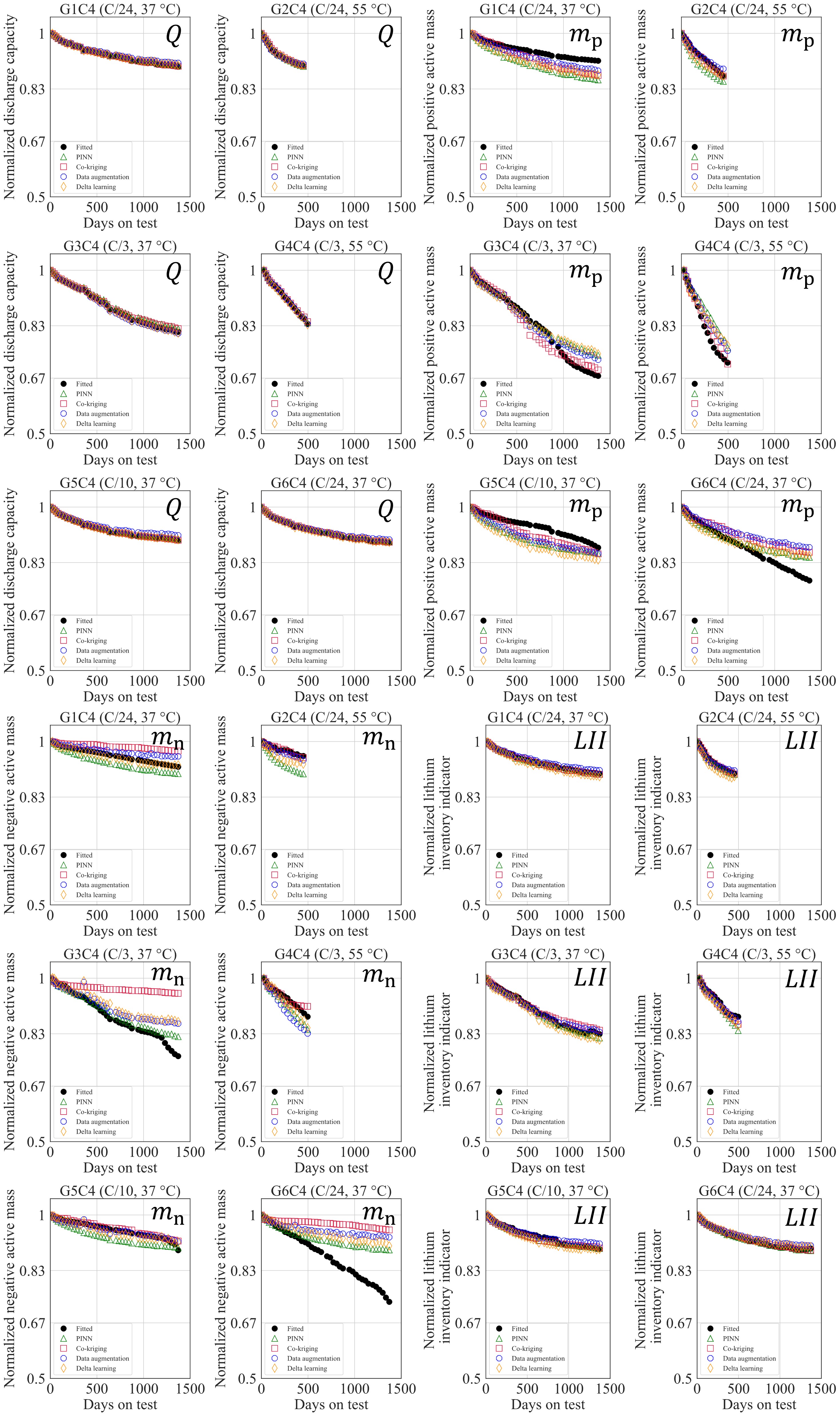}
\centering
\caption{Trajectory prediction results for the forth fold (forth cell of each group).}
\label{fig:Trajectory_4}
\end{figure}

 \clearpage
\subsection*{A5. Half-cell surrogate model}

\begin{figure} [h]
\includegraphics[width=0.9\textwidth]{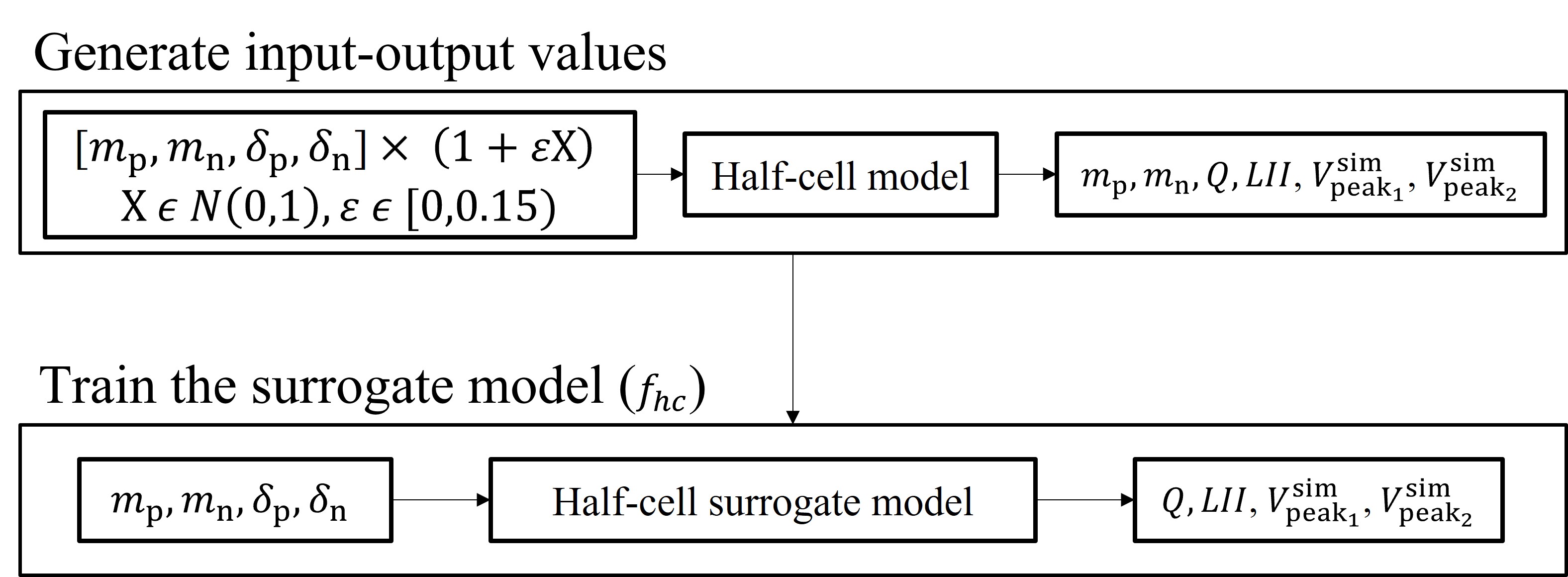}
\centering
\caption{Training half-cell surrogate model.}
\label{fig:hc_surrogate}
\end{figure}

As described in Sec. \ref{subsec:PINN}, we integrate a half-cell surrogate model into the loss function of the PINN model to ensure both differentiability and consistency of the computation graph between the inputs and outputs. Here, we provide detailed information about how we implemented the half-cell surrogate model for this purpose.

As depicted in Fig. \ref{fig:hc_surrogate}, we initially utilize the half-cell model to obtain health parameters and two simulated $dQ/dV (V)$ peak positions(voltage values) by inputting perturbed values of the half-cell model parameters ($\delta_{\mathrm{p}}, \delta_{\mathrm{n}}, m_{\mathrm{p}}, m_{\mathrm{n}}$). The objective is to encompass a broader range of possible parameters for inputting into the half-cell model. This is necessary because in the loss function, we need to be able to map the deviated values of half-cell parameters, which are predicted by the network, to health parameters.

The perturbed values are obtained through standard normal sampling from a 15\% range of perturbation for each parameter. Subsequently, we use the true values, perturbed values, and corresponding degradation and peak position values as input and output values, respectively. Using these values, we establish a straightforward mapping function between the half-cell model parameters and capacity, lithium inventory indicator, and two $dQ/dV (V)$ curves peak voltage values ($Q$, $LII$, $\mathrm{V}_{\mathrm{peak}_1}^{\mathrm{sim}}$, $\mathrm{V}_{\mathrm{peak}_2}^{\mathrm{sim}}$) to train a surrogate neural network ($f_{hc}$) with perfect accuracy in mapping these values.

\end{document}